\documentclass[aps,prb,amsfonts, twocolumn,longbibliography, superscriptaddress%,onecolumn,notitlepage%,footinbib%,11pt
amsmath,amssymb,dvipsnames]{revtex4-1}
\usepackage{hyperref}
\hypersetup{colorlinks=true, linkcolor=blue, citecolor=red, urlcolor=blue}
\usepackage{tikz}
\usetikzlibrary{arrows}
\usetikzlibrary{decorations.pathmorphing}
\usepackage{hyperref}
\usepackage{amssymb}
\usepackage{amsmath}
\usepackage{graphicx}
\usepackage{soul}
\usepackage{mathtools}
\usepackage{siunitx}

\newcommand{\comment}[1]{}

\newcommand{\RN}[1]{%
\textup{\uppercase\expandafter{\romannumeral#1}}%
}
\newcommand{\intrakk}{V^{\text{intra}}_{\bold{k}\bold{k}'}}
\newcommand{\interkk}{V^{\text{inter}}_{\bold{k}\bold{k}'}}
\newcommand{\Uintrakk}{U^{\text{intra}}_{\bold{k}\bold{k}'}}
\newcommand{\Uinterkk}{U^{\text{inter}}_{\bold{k}\bold{k}'}}
\newcommand{\Uintra}{U_{\text{intra}}}
\newcommand{\Uinter}{U_{\text{inter}}}
\newcommand{\UattrI}{U_{\text{attr}}}
\newcommand{\dif}{\mathrm{d}}

\newcommand{\DE}[1]{\tilde{\xi}_{#1}} %\newcommand{\DE}[1]{E_{#1}}

\newcommand{\DET}[1]{E_{#1}} %\newcommand{\DET}[1]{\tilde{E}_{#1}}
\newcommand{\DXi}[1]{\eta_{#1}} %\newcommand{\DXi}[1]{\tilde{\xi}_{#1}}
\newcommand{\DA}{Y} %\newcommand{\DA}{A}
\newcommand{\DB}{Z} %\newcommand{\DB}{B}

\newcommand{\Da}[1]{\kappa_{#1}} %\newcommand{\Da}[1]{a_{#1}}
\newcommand{\DR}{R}
\newcommand{\DS}{S}
\newcommand{\DT}{T}

\newcommand{\DxiT}[1]{\tilde{E}_{#1}} %\newcommand{\DxiT}[1]{\tilde{\xi}_{#1}}
\newcommand{\Dp}[1]{\lambda_{#1}} %\newcommand{\Dp}[1]{p_{#1}}
\newcommand{\DU}[1]{u_{#1}}
\newcommand{\DV}[1]{v_{#1}}
\newcommand{\DM}{\bold{W}} 

\newcommand{\Dg}[1]{g_{#1}} %becomes 1 for B=0
\newcommand{\Dh}[1]{h_{#1}} %becomes 0 for B=0

\usetikzlibrary{patterns}

\begin{document}

\title{Two bands Ising superconductivity from Coulomb interactions in monolayer $\text{NbSe}_2$}

\author{Sebastian H\"orhold}
\affiliation{Institute for Theoretical Physics, University of
Regensburg, 93040 Regensburg, Germany}
\author{Juliane Graf}
\affiliation{Institute for Theoretical Physics, University of
	Regensburg, 93040 Regensburg, Germany}
\author{Magdalena Marganska}
\affiliation{Institute for Theoretical Physics, University of
	Regensburg, 93040 Regensburg, Germany}
\author{Milena Grifoni}
\affiliation{Institute for Theoretical Physics, University of
Regensburg, 93040 Regensburg, Germany}
\date{\today}

\begin{abstract}
{The nature of superconductivity in monolayer transition metal dichalcogenides is still an object of debate. It has already been argued  that  repulsive  Coulomb interactions,  combined with the disjoint Fermi surfaces around the $K$, $K'$ valleys  and at the $\Gamma$ point, can lead to superconducting instabilities in monolayer $\text{NbSe}_2$. Here,  we demonstrate the two bands nature of superconductivity in $\text{NbSe}_2$.  It arises from the competition of repulsive long range intravalley and short range intervalley interactions together with Ising spin-orbit coupling.
	 The two distinct superconducting gaps, one for each spin-orbit split band, consist of a mixture of $s$-wave and $f$-wave components. 
	 Their different amplitudes are due to different normal densities of states of the two bands at the Fermi level. 
Using a microscopic multiband BCS approach, we derive and self-consistently solve the gap equation, demonstrating the stability of nontrivial solutions in a realistic parameter range. We find a universal behavior of the temperature dependence of the gaps and of the  critical in-plane field   which is  consistent with various sets of existing experimental data. 
}
\end{abstract}
\maketitle

%\tableofcontents

%\magda{TO DO:\\
%- prove that our $\Delta$'s yield a minimum of the free energy, not just an extremum\\
%- consider the role of plasmons in screening and perhaps setting the cutoff $\Lambda$\\
%- identify explicitly singlet and triplet contributions to the pairing\\}

\section{Introduction}
\label{intro}
In conventional materials the dominance of repulsive Coulomb interactions is in general detrimental to superconductivity. Nevertheless, it has long been known that, accounting for long range oscillatory contributions in some fermionic systems, superconductivity can still arise by the so-called Kohn-Luttinger mechanism\cite{Kohn:prl1965}. 
It is also well recognized that Coulomb interactions are strongly enhanced in layered systems like the cuprates or iron-pnictides \cite{Mazin:nature2010}, and that they might be at the origin of superconductivity in twisted bilayer graphene and other novel two-dimensional materials \cite{Qui:advmat2021}.

In this context, unconventional superconductivity in   two-dimensional transition metal dichalcogenides (TMDCs), systems with fragmented Fermi surface, has attracted much attention in recent years. With focus on the observation of superconductivity in heavily doped molybdenum disulfide ($\text{MoS}_2$) \cite{Taniguchi:apl2012,Ye:science2012,Saito:natphys2016},
Rold\'an et al.~\cite{Roldan:prb2013} have 
suggested that the competition between short and long range processes, both of them repulsive, can lead to an effective attraction resulting in superconducting pairing. 
Later theoretical works have further focused on  
various scenarios for possible mechanisms of superconductivity and non-trivial topological phases in this system  \cite{Yuan:prl2014,Hsu:natcomm2017,Oiwa:prb2018}.

While MoS$_2$ becomes superconducting after doping,  monolayer NbSe$_2$ is an intrisic van der Waals superconductor. Due to the large Ising spin-orbit coupling, locking Cooper pairs out-of-plane, it  exhibits critical in-plane magnetic fields well above the Pauli limit\cite{Xi:natphys2016}. 
Recently,  Shaffer et al. 
\cite{Shaffer:prb2020} have proposed a detailed phase diagram of possible unconventional  superconducting phases of monolayer NbSe$_2$ upon  application of an in-plane magnetic field and with the  addition of Rashba spin-orbit coupling.
%\replace{, while mirage gaps in finite magnetic fields have been discussed in}{ . 
The presence  of magnetic field applied in a direction perpendicular to the spin-orbit fields is also thought to cause the formation of equal-spin triplet pairs in TMDCs with natural singlet pairing \cite{Zhou:prb2016,Ilic:prl2017,Moeckli:prb2018,Tang:prl2021}. 

Despite the many predictions of exotic phases  by tuning  doping or  magnetic fields, little theoretical attention has been put on the \textit{intrinsic two-bands character} of the superconductivity in monolayer TMDCs, the topic of this work. It arises from the  large  
Ising spin-orbit coupling in combination with short and long range Coulomb repulsion.  
 We focus on monolayer $\text{NbSe}_2$, but the ideas exposed in this work are rather general and can be applied to characterize superconductivity in other van der Waals materials,  or in systems with two disjoint (also spin-split) Fermi surfaces, in presence of competing interactions -- repulsive, attractive, or a mixture of both. \\
 Specifically, we expand the original idea of Rold\'an et al.~\cite{Roldan:prb2013} to include the effects of Ising spin-orbit coupling and later also  of an in-plane magnetic field on the superconducting phase transition.
 Starting from  repulsive interactions and disjoint Fermi surfaces around the $K$ and $K'$ points in $\text{NbSe}_2$, we find two distinct superconducting gaps, one for each spin-orbit split band, both consisting of a mixture of $s$-wave and $f$-wave components. Using a microscopic multiband BCS approach we derive and self-consistently solve the coupled gap equations, demonstrating the stability of nontrivial solutions in a realistic parameter range.
Similar to standard single band BCS, we find a universal behavior of the mean gap vs. temperature.

To date, the possibility to 
 produce high-quality monocrystals with few or even one single layer by mechanical exfoliation or molecular beam epitaxy  \cite{Xi:natphys2016,Ugeda:natphys2016,Xing:nanolett2017,Khestanova:nanolett2018,Hamill:natphys2021,Kuzmanovic:arxiv2021}, makes it possible to get access to   
 the pairing  mechanism, and  to some of the universal features of superconductivity in monolayer NbSe$_2$  discussed in this work.  For example, 
the temperature dependence of the gaps and of the  critical in-plane field  are consistent with various sets of existing experimental data  \cite{Xing:nanolett2017,Zhao:natphys2019,Wan:arxiv2021}. The presence of two gaps is further in agreement with the recent observation of a collective Leggett mode \cite{Wan:arxiv2021}.
%  Finally,  our results in finite magnetic field reveal a  triplet component of the order parameter. \magda{We have not analyzed the spin composition of our $\Delta$'s, so this claim is not well supported. -Magda} This finding supports the possible experimental evidence of triplet superconductivity in bilayer NbSe$_2$ according to the work of  Kuzmanovi\'{c} et al. \cite{Kuzmanovic:arxiv2021}. \magda{As in the sentence before. -Magda}

%
\indent The paper is structured as follows. 
In Sec.~\ref{Sec:Model} we briefly recall the band structure of $\text{NbSe}_2$ and present a minimal low energy model which captures the main features 
around the Fermi energy. In Sec.~\ref{Sec:two-gaps} the coupled gap equations are obtained and the predicted temperature dependence of the gaps is investigated for two parameter sets. The impact of an in-plane magnetic field and the dependence of the critical field on temperature are  discussed in  Sec.~\ref{Sec:magnetic-field}.
Finally, conclusions are drawn in Sec.~\ref{Sec:Conclusions}.
Some of the detailed derivations are deferred to the appendix. \\

\section{Band structure and minimal model for monolayer $\text{NbSe}_2$} 
\label{Sec:Model}
\begin{figure}[ht]
\begin{center}
\includegraphics[width=\columnwidth,angle=0]{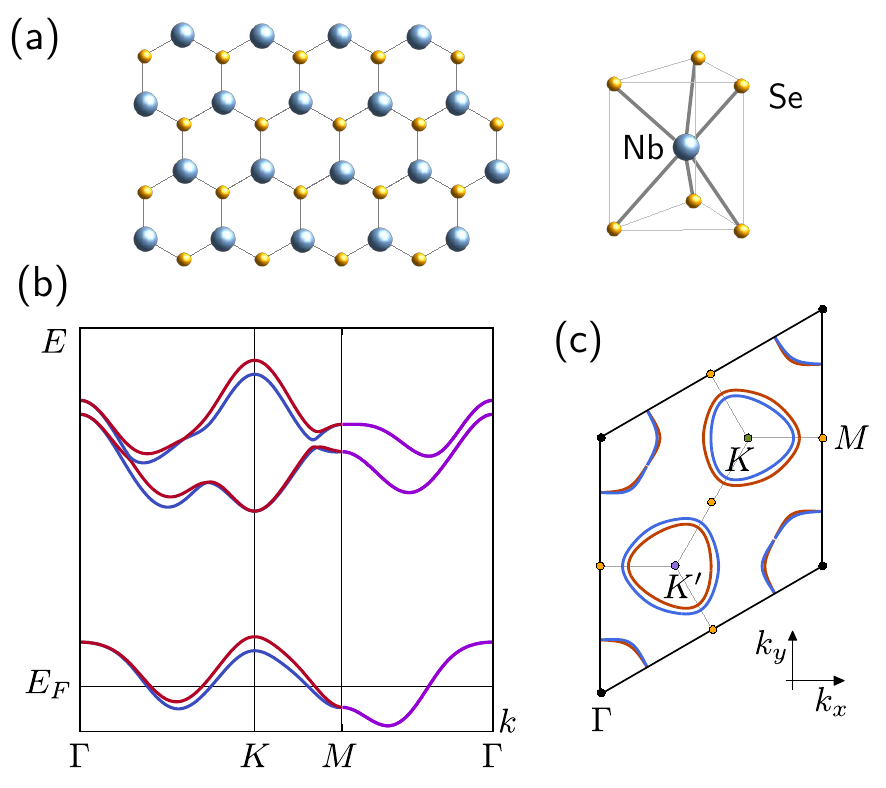}
\caption{\small{Lattice, band structure and Brillouin zone of NbSe$_2$. (a) Top view of the lattice and the trigonal prismatic unit cell. (b) The six spin-resolved bands  of NbSe$_2$  closest to the  Fermi energy $E_F$. They have been obtained within a tight-binding model involving  three $d$ orbitals of the Nb atoms. The Ising spin-orbit coupling induces a large splitting of the valence bands at the $K$ and $K'$ points. (c) The Brillouin zone of NbSe$_2$, with spin-resolved Fermi surface  pockets near the $K,K'$ and $\Gamma$ points. }}
\label{fig:lattice-bands}
\end{center}
\end{figure}

%
% Please start from a brief description on monolayer TMDs, the effect of broken inversion symmetry leading to Ising SOC. Tight binding model from which a minimal model can be obtained near the K and K' valleys and comparison with the prediction of the minimal model.\\
% \\
Monolayer transition metal dichalcogenides $MX_2$ are made up of a single layer of $M$ transition metal atoms  sandwiched between two layers of $X$ chalcogen atoms. The metal and chalcogen atoms can enter in various combinations, which make them very attractive for applications \cite{Geim:nature2013};  superconductivity has been largely  investigated in TMDCs with  $M$ = Mo, Nb and $X$ = S, Se \cite{Taniguchi:apl2012,Ye:science2012,Saito:natphys2016,Xi:natphys2016,Xi:prl2016}.  As shown in Fig.~\ref{fig:lattice-bands}(a), each $M$ atom binds to the six nearest $X$ atoms that together form the trigonal prismatic unit cell of the lattice. Projecting these layers onto a plane yields a honeycomb lattice similar to the one found in graphene. 
The primitive unit cell of the $M$ sublattice has the area  $\Omega = \frac{\sqrt{3}}{2}a^2 = 10.28\, {\si{\angstrom}}^2$ with the lattice constant $a=3.445\, \si{\angstrom}$.
The dispersion relation of monolayer NbSe$_2$ along high symmetry lines is shown in Fig.~\ref{fig:lattice-bands}(b). 
It has been obtained within a tight-binding (TB) model where only the three orbitals $d_{z^2}$, $d_{x^2-y^2}$ and $d_{xy}$  of the metal atom are retained \cite{Liu:prb2013,Kim:prb2017,He:commphys2018}, with the TB parameters for NbSe$_2$ taken from Ref.~[\onlinecite{He:commphys2018}]. The strong atomic spin-orbit coupling (SOC) due to the heavy transition metal $M$ is included in the band structure calculation. 
 Since the lattice shown in Fig. \ref{fig:lattice-bands}(a) possesses an out-of-plane mirror symmetry, the crystal field is restricted to the in-plane direction of the system. Taking into account that the electronic motion is confined to the 2D lattice, the effective SOC field felt by the moving charges  also points in the out-of-plane direction. Consequently, the electron spin is also quantized along this axis and remains a good quantum number.\cite{Saito:natphys2016} This kind of SOC is known as   Ising spin-orbit coupling \cite{Xi:natphys2016}. Its effect is to remove spin degeneracy of the bands by inducing a momentum dependent energy shift. The latter is very prominent in the valence bands near the $+K$ and $-K$ points (or simply $K$ and $K'$) related by time-reversal symmetry. This is due to the fact that the $d$-bands there are predominantly given by the linear combinations $d_{x^2-y^2}\pm id_{xy}$  with angular momentum  $L=\pm 2 \hbar$. Along the high symmetry $\Gamma M$ line the valence band is spin degenerate. 

When viewed within the rhomboidal Brillouin zone, the fragmentation of the Fermi surface of NbSe$_2$, which will be crucial in our discussion of unconventional superconductivity, becomes apparent. As depicted in Fig.~\ref{fig:lattice-bands}(c), the Fermi surface is composed of hole pockets around the $K$ and  $K'$ valleys and the $\Gamma$ point. The spin-resolved pockets around  $K$ and $K'$ display a trigonal warping and are related by time-reversal symmetry. 
%give rise to an additional valley degree of freedom for charge carriers like holes and electrons.
\\
 \subsection{A low energy minimal model for NbSe$_2$} 
Superconductivity is a low energy phenomenon originating from the binding of electrons residing close to the Fermi energy.  Hence, in the following  we will only consider the valence bands and will focus on the features close to the Fermi energy. 
Furthermore, as the mechanism we shall discuss strongly relies on the existence of disconnected Fermi surfaces related by time reversal symmetry, we shall focus on the dispersion around the $K$ and $K'$ valleys and disregard the contribution of the $\Gamma$ Fermi surface. The latter can admit at most $s$-wave pairing \cite{Shaffer:prb2020} and is not relevant for the mechanism discussed here, leading to a dominant  $f$-wave channel.
%
%\begin{figure}[ht]
%\begin{center}
%\includegraphics[width=\columnwidth,angle=0]{figures/minimal-model-illustration.pdf}
%\caption{\small{Band structure of NbSe$_2$ along the $k_y=0$ axis, crossing some of the high symmetry points. Concrete parameters where taken from Kim et al.}}
%\label{fig:lattice-bands}
%\end{center}
%\end{figure}
%

Our  aim is to develop a minimal low energy  model for superconductivity in NbSe$_2$. Hence, instead of using the full tight-binding models mentioned above, we  restrict the following discussion to a hyperbolic fit to the dispersion in the two valleys. The fitting parameters are obtained from two tight-binding parametrizations \cite{Kim:prb2017,He:commphys2018}, as discussed below. For simplicity, the trigonal warping far from the Dirac points is neglected. Then the  hyperbolic dispersion for a particle of spin $\sigma$   and momentum $\bold{k}$ measured from the Dirac point  $\tau K$, with $\tau=\pm$, is written as
\begin{equation}
    \varepsilon_{\tau\sigma}(\bold{k}) = {\epsilon}_{\tau\sigma}^0+m_{\tau\sigma} - \sqrt{(\hbar v_{F,\tau\sigma})^2\bold{k}^2 + m_{\tau\sigma}^2}\,.
    \label{eq:hyperbolic_dispersion}
\end{equation}
In the above equation $v_{F,\tau\sigma}$ is the Fermi velocity, $m_{\tau\sigma}$ is a mass-like parameter and 
%$\Tilde{\epsilon}_i = \epsilon_i^{u}+m_i$, with
$\epsilon_{\tau\sigma}^{0}$ the upper limit of the band, i.e. the energy directly at the $\tau K$ point. 
 Since time-reversal symmetry is preserved by the SOC, it holds
$\varepsilon_{\tau \sigma}(\bold{k})= \varepsilon_{-\tau -\sigma}(-\bold{k}):=\varepsilon_{\bar\tau \bar\sigma}(\bar{\bold{k}})$, where  we used the shorthand notation $-\bold{k}:=\bar{\bold{k}}$, $-\tau:=\bar\tau$ and  $-\sigma:=\bar\sigma$. 
This symmetry allows us to restrict our considerations to the $K$ valley by  introducing the pseudospin indices 
\begin{equation}
    i = \left\{ 
    \begin{matrix}
    1 & \ \text{for} & (K, \uparrow) \\
    2 & \ \text{for} &  (K, \downarrow) 
    \end{matrix}
     \right.\,, \quad 
    \bar{i} = \left\{ 
    \begin{matrix}
     \bar{1} & \ \text{for} &   (K', \downarrow)\\
  \bar{2} & \ \text{for} & (K', \uparrow)
    \end{matrix}
    \right. .
\end{equation}
Here, $i=1 \, (2)$ denotes the upper (lower) band in the $K$ valley, while $\bar{i}=\bar{1}\, (\bar{2})$ are the time-reversed upper (lower) bands in the $K'$ valley. 
%Thus, we can conveniently focus on the valley $K$ with $\tau=+$. In this case spin $\sigma =\uparrow (\downarrow)= + (-)$
%refer to the upper (lower) bands. 
With this notation  the energy relative to the chemical potential $\mu$ in the $K$ valley can be written as 
\begin{equation}
    \xi_i(\bold{k}) = \varepsilon_i(\bold{k}) - \mu = {\xi}_i^0+m_i - \sqrt{(\hbar v_{F,i})^2\bold{k}^2 + m_i^2} \,,
    \label{eq:hyperbolic_dispersion_mu}
\end{equation}
where  $\xi_i^{0} = \epsilon_i^{0}-\mu$.
%in $\Tilde{\xi}_i$. %
%
\begin{figure}[ht]
    \begin{center}
        \includegraphics[width=\columnwidth,angle=0]{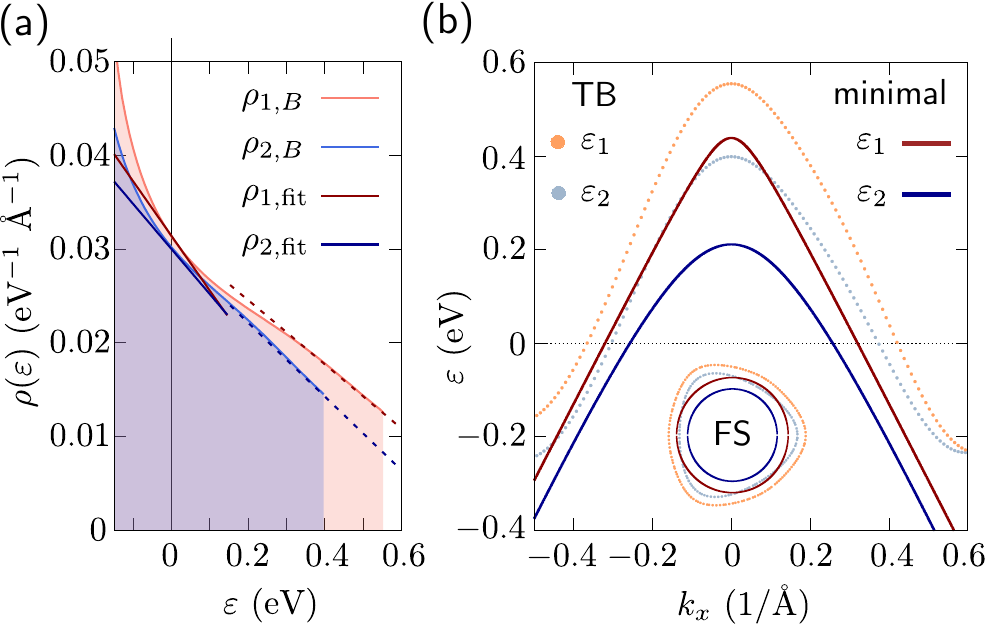}
        \caption{Density of states (DOS) and energy bands in the minimal model. (a) DOS in the $K$ valley associated to the spin-orbit split valence bands. The coloured regions depict the DOS  $\rho_{1,2,\textnormal{TB}}$  obtained numerically from the tight-binding model. The solid lines are the linear fits  $\rho_{1,2,
        \textnormal{fit}}$ to those DOS close to the Fermi level, used to estimate the parameters of the effective model. Dashed lines mark the linear fits to the DOS higher on the Dirac cone, for comparison. (b) The tight-binding and the effective model dispersion and (in the inset) the Fermi surfaces in the $K$ valley, with $k_x$ measured with respect to $K$. Solid lines show results for the effective model, the dotted curves display the numerical tight-binding bands with the parametrization in \cite{Kim:prb2017}. }
        \label{fig:dos-bands}
    \end{center}
\end{figure}
%
% Here is an example for a split equation
% \begin{equation}
% \label{eq:dispersion}
% \begin{aligned}
% \hat{H}=&\sum_{\alpha k \sigma}\epsilon_{\alpha k}c^{\dag}_{\alpha k \sigma}c_{\alpha k \sigma}\\
% &+\sum_{i \alpha k \sigma}\left[ {\rm t}_{i\alpha k \sigma} a_{i}^{\dag}c_{\alpha k \sigma}+ {\rm t}^{*}_{i\alpha k \sigma} c^{\dag}_{\alpha k \sigma}a_{i}\right]\;.
% \end{aligned}
% \end{equation}
% A comparison between the tight-binding bands and the dispersion in Eq. (\ref{eq:dispersion}) is shown in  \ref{fig:lattice-bands}. 
%
\subsection{Density of states}
%Numerical results for the  density of states are shown in Fig.  \ref{fig:DOS}
Making use of the approximate band structure Eq.~\eqref{eq:hyperbolic_dispersion_mu}, we can  get to an expression for the band resolved density of states (DOS) per unit area at the $K$-Dirac cone. %Its calculation is straightforward after transforming the sum over momenta into an integral over the two dimensional k-space.
We find
\begin{equation}
\label{eq:DOS}
    \rho_i(\varepsilon) := \frac{1}{N\Omega} \sum_{\bold{k}} \delta( \varepsilon -\varepsilon_i(\bold{k}) )%
    = \left( \rho_{Fi}-d_{i} \varepsilon \right) \theta(\epsilon_i^0-\varepsilon)\,,
\end{equation}
where $N$ is the number of Nb atoms in the lattice, with the factor of the linear term $d_{i} = 1/(2\pi(\hbar v_{F,i})^2)$ and the constant term $\rho_{Fi} = ({\epsilon}_i^0+m_i)  d_{i}:=\Tilde{\epsilon_i}d_i$. Hence, 
the DOS directly at the chemical potential is given by 
\begin{equation}
    \rho_i(\mu) = \left( \rho_{Fi}-d_{i} \mu \right)\, \theta(\epsilon_i^0-\mu) = 
    ({\xi}_i^0+m_i)d_i \;\theta(\xi_i^0)\,.
    \label{eq:DOS_i}
\end{equation}
We have two guiding principles in constructing the minimal model: \\
(i) The DOS for each band must be the same as in the tight-binding model within the relevant energy range around the Fermi level.\\
(ii) The spin-orbit splitting between the two bands at the Fermi level should be correct (this will be important when we consider the evolution of the gap in magnetic field).\\
In the following we set the zero of the energy at the Fermi level of the normal system. We can fix the free parameters $\Tilde\epsilon_i$ and $v_{F,i}$ of the minimal model by requiring that the DOS  in Eq.~(\ref{eq:DOS}) assumes the value $\rho_{Fi}$ at the Fermi level and its slope is determined by $d_i$. One of the masses is chosen arbitrarily to be $m_1 = 0.1$~eV. The value of $m_2$ (and hence also the parameter $\epsilon_2^0=\Tilde\epsilon_2 -m_2$) is set by fixing the values of the spin orbit splitting at the Fermi level.
Explicitly, we require
\begin{equation}
    \Delta_{\text{SOC}}:= \varepsilon_1(k_{F,1}) -\varepsilon_2(k_{F,1})= -\varepsilon_2(k_{F,1})\,,
    \label{eq:SOC}
\end{equation}
where $k_{F,i}$ are the Fermi momenta of the two bands obtained in full tight-binding, satisfying $\varepsilon_i(k_{F,i})=0$. The value of $\Delta_{\text{SOC}}$ is taken from the tight-binding calculation as an average between the spin-orbit splitting in the $\Gamma K$ and in the $KM$ direction.
The results of the fitting procedure are illustrated in Fig.~\ref{fig:dos-bands} where the tight-binding bands and DOS were calculated using the parameter set given in [\onlinecite{He:commphys2018}], denoted in the following as set $B$. An alternative set\cite{Kim:prb2017} (plots not shown) is denoted as set $A$. The resulting minimal model parameters are shown in Table~\ref{table:parameters}.
%
%\begin{figure}[ht]
%	\begin{center}
%		\includegraphics[width=8cm,angle=0]{figures/DOS_at_K.png}
%		\caption{\small{Density of states of NbSe2. Numerical results. }}
%		\label{fig:DOS}
%	\end{center}
%\end{figure}
%
\begin{table}[h!]
\centering
\begin{tabular}{c c | c c c c c c}
set & $i$ & $\rho_{Fi}$ & $d_{i}$ & $\epsilon_i^0$ & $\tilde{\epsilon}_i$ & $\hbar v_{F,i}$ & $m_i$\\ 
 \hline\hline
 $A$\cite{Kim:prb2017} & 1  & 0.0385 & 0.09 & 0.328 & 0.428 & 1.33 & 0.1 \\ 
 & 2  & 0.046   & 0.13 & 0.199  & 0.354 & 1.106 & 0.155 \\
 \hline
$B$\cite{He:commphys2018} & 1  & 0.0314 & 0.0583 & 0.439 & 0.539 & 1.652 & 0.1 \\ 
 & 2  & 0.03   & 0.0481 & 0.209  & 0.624 & 1.819 & 0.415 \\
\hline\hline
\end{tabular} 
\caption{
\label{table:parameters}
Parameters from the linear fit to the DOS and the SOC splitting at the Fermi energy. %
%The first three columns contain the numerical fit parameters from \cite{Marganska} and the last three are the corresponding parameters for the hyperbolic dispersion, calculated by using the definition given below \eq (\ref{chap1:DoS}). 
Units for $\rho_{Fi}$, $d_{i}$ and $\hbar v_{F,i}$ are ${\text{eV}}^{-1}{\si{\angstrom}}^{-2}$, ${\text{eV}}^{-2}{\si{\angstrom}}^{-2}$ and eV{\AA}, respectively. The values of spin-orbit splitting at the Fermi level are $\Delta_{\text{SOC}}^A = 0.0253$~eV and $\Delta_{\text{SOC}}^B = 0.092$~eV.}
%The remaining quantities have the same units as in Table \ref{chap1:parameterTable}.}
%\label{chap1:tableParameterDOS}
%$\frac{1}{\text{eV}\si{\angstrom}^2}$ and $\frac{1}{\text{eV}^2\si{\angstrom}^2}$
\end{table}
\section{Two-bands superconductivity and coupled gap equations} 
\label{Sec:two-gaps}
We now turn to  the mechanism inducing the two-bands  superconducting phase in NbSe$_2$. For this purpose we will focus on the two partially occupied bands around the Fermi level which give rise to the spin separated Fermi surfaces at the $K$ and $K'$ points discussed in the previous section. Another smaller pocket might be found at the $\Gamma$ point; however it will not be part of the further discussion since it is expected to only play a subordinate role \cite{Shaffer:prb2020}.  Instead of the conventional pairing mechanism that leads to the formation of Cooper pairs, i.e. the phonon-mediated attraction of two electrons, we will now consider the Coulomb repulsion of such electrons and hence an unconventional pairing.  While phonons are believed to contribute  the dominant mechanism for bulk NbSe$_2$ \cite{Valla:prl2004,Noat:prb2015,Heil:prl2017,Sanna:npj2022}, this should not be the case for the monolayer case\cite{Wan:arxiv2021}.  Further, Coulomb interactions are not well screened in a monolayer, and hence short range and long contributions should be included.

\begin{figure}[ht]
	\begin{center}
		\includegraphics[width=\columnwidth,angle=0]{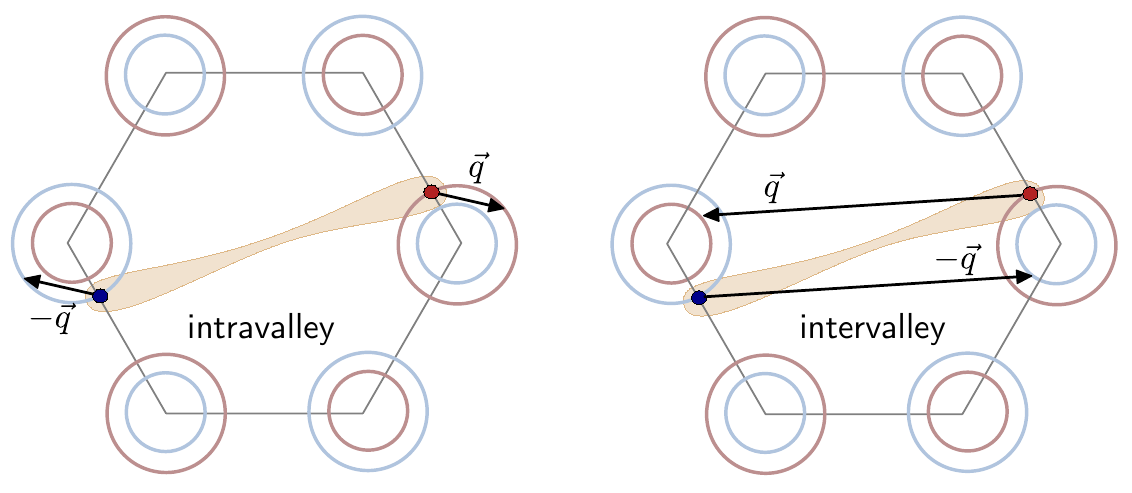}
		\caption{\small{Superconductivity in monolayer TMDCs can arise from the interplay of repulsive long and short range interactions. On the left, a typical intravalley scattering process which involves small momentum transfer $\pm \vec{q}$ is shown. On the right, intervalley scattering with large momentum transfer $\pm \vec{q}$ is depicted.  The scattered Cooper pairs are related by time reversal and hence reside in different valleys.    }}
		\label{fig:interactions}
	\end{center}
\end{figure}

According to these considerations, we start from an interacting Hamiltonian with  conventional, spin independent,  Coulomb interaction. By retaining only scattering processes among time-reversal related Cooper pairs, the total Hamiltonian is the sum of a single particle and an  interaction part,
\begin{equation}
    \hat{H}_{\text{tot}}= \hat{H}_{\text{sp}} + \hat{H}_{\text{int}}\,.
\end{equation}
The single particle contribution follows from the minimal model from the previous section and  reads 
\begin{equation}
   \hat{H}_{\text{sp}} = \sum_{\sigma\tau\bold{k}} \varepsilon_{\tau \sigma}(\bold{k}) \hat{c}^\dagger_{\bold{k}\tau\sigma} \hat{c}_{\bold{k}\tau\sigma} \,,
\end{equation}
with the dispersion provided by Eq. (\ref{eq:hyperbolic_dispersion}).
%and where $\sigma$ and $\tau$ account for the spin and valley degrees of freedom of the electrons. 
The interaction Hamiltonian \cite{Roldan:prb2013} 
\begin{align}
\label{eq:interaction_ham}
    \hat{H}_{\text{int}} = \frac{1}{2} \sum_{\sigma\tau\bold{k}\bold{k}'} \big[ %
    &\intrakk \hat{c}^\dagger_{\bold{k}\tau\sigma}
             \hat{c}^\dagger_{\bar{\bold{k}}\bar\tau \bar\sigma}
             \hat{c}_{\bar{\bold{k}}' \bar\tau \bar\sigma}
             \hat{c}_{\bold{k}'\tau\sigma} \notag\\ 
   + & \interkk \hat{c}^\dagger_{\bar{\bold{k}} \bar\tau\sigma}
             \hat{c}^\dagger_{\bold{k}\tau \bar\sigma}
             \hat{c}_{\bar{\bold{k}}' \bar\tau \bar\sigma}
             \hat{c}_{\bold{k}'\tau\sigma}         
    \big]
\end{align}
accounts for both intravalley and  intervalley scattering processes. 
For an electron with momentum $\bold{k}$ and spin $\sigma$ that is located at the $K$ valley, its time-reversal partner will be located at $K'$ with opposite momentum.  Thus, as shown in Fig. \ref{fig:interactions}, there are now two possible scattering mechanisms, mediated by $\intrakk$ and $\interkk$, that can occur between the members of a  pair.
For an intravalley process the scattered electrons stay within  their initial valley in $\bold{k}$-space, i.e. the valley index will be conserved. This means that the exchanged momentum $\bold{q=k-k'}$ is small, which corresponds to a long-ranged interaction in real space. The intervalley scattering on the other hand describes a short ranged Coulomb interaction with a large exchanged momentum of the order of $2|\bold{K}|$. In this process the electrons swap their valley and hence  $\tau$ flips its sign. Note that since the Coulomb interaction conserves spin, the intervalley scattering transforms a Cooper pair residing on the inner (outer) Fermi surface into a pair residing on the outer (inner) Fermi surface. This process thus couples the two condensates.
 The two sums in Eq. (\ref{eq:interaction_ham}) run over a shell around the Fermi surface whose thickness will be denoted as $\Lambda$. In fact, as in the conventional BCS theory, the restriction of the sums in momentum space to time-reversal partners is appropriate  at low energies where only  electrons in the vicinity of the Fermi energy are involved \cite{Bardeen:physrev1957}. For a phonon-mediated interaction the shell  thickness is in the order of the Debye energy $\hbar \omega_{\text{D}}$. 
 %\magda{For the pairing mediated by competing repulsion processes the resulting attraction is only effective, not associated with the presence of a mediating excitation such as a phonon. Hence, there is no obvious energy scale for $\Lambda$.} 
 Here we shall  assume a shell thickness also in the meV range. 
 \subsection{Mean field Hamiltonian}
 \label{subsec:Mean_field}
 %
 %Up to this point is not obvious that the interplay of the two interactions can result in an effective attraction. 
 Due to the complexity of the scattering processes described by the interaction Hamiltonian Eq. (\ref{eq:interaction_ham}), 
 we 
 simplify the problem by performing a mean field approximation \cite{Bardeen:physrev1957} on both interaction terms. By introducing the pairing functions 
\begin{align}
    \Delta_{\tau\sigma}^{\text{intra}}(\bold{k})&= - \sum_{\bold{k}'} 
    \intrakk \langle \hat{c}_{\bar{\bold{k}}'\bar\tau \bar\sigma}
                     \hat{c}_{\bold{k}'\tau\sigma} \rangle \notag \,, \\ 
\Delta_{\tau\sigma}^{\text{inter}}(\bold{k}) & =    -\sum_{\bold{k}'} \interkk \langle \hat{c}_{\bar{\bold{k}}'\bar\tau\sigma}
                     \hat{c}_{\bold{k}'\tau\bar\sigma} \rangle\,,
    \label{eq:gap_intra_inter}
\end{align} and    by making use of the fermionic anticommutation relations, we can express the interaction Hamiltonian in the compact form 
%we obtain the mean field interaction Hamiltonian
%\begin{equation}
 %   \hat{H}_{\text{int}}^{\text{mf}} = -\frac{1}{2} \sum_{\bold{k}\tau\sigma} \big[  \Delta_{\tau,\sigma}^{\text{intra}}(\bold{k})
  %  \hat{c}^\dagger_{\bold{k}\tau\sigma}
   % \hat{c}^\dagger_{\bar{\bold{k}}\bar\tau \bar\sigma} \notag 
   % - \Delta_{\tau,\sigma}^{\text{inter}}(\bold{k})
    %\hat{c}^\dagger_{\bold{k} \tau \bar\sigma}
    %\hat{c}^\dagger_{\bar{\bold{k}} \bar\tau\sigma}
    %\big]
    %\label{eq:MFInteraction}
%\end{equation}
%
%
\begin{equation}
    \hat{H}_{\text{int}}^{\text{mf}} = -\frac{1}{2} \sum_{\bold{k}\tau\sigma} \big[  \Delta_{\tau\sigma}(\bold{k})
    \hat{c}^\dagger_{\bold{k}\tau\sigma}
    \hat{c}^\dagger_{\bar{\bold{k}}\bar\tau \bar\sigma} \notag 
    +  (\Delta_{\tau\sigma}(\bold{k}))^*
    \hat{c}_{\bar{\bold{k}} \bar\tau \bar\sigma}
    \hat{c}_{\bold{k}\tau\sigma}
    \big] \,,
    \label{eq:MFInteractionTerm}
\end{equation}
with  the global gaps 
\begin{align}
    \Delta_{\tau\sigma}(\bold{k}) =  \Delta_{\tau\sigma}^{\text{intra}} (\bold{k})-  \Delta_{\tau\sigma}^{\text{inter}}(\bold{k})\,.
    \label{eq:definition_gaps}
\end{align}
In Eq. (\ref{eq:MFInteractionTerm})  irrelevant constant terms, encapsulated in an overall contribution $C$, have been omitted. 
%obtained from the mean-field approximation and the spin and valley dependent superconducting gap functions $\Delta_{\tau,\sigma}(\bold{k})$ are defined as
%
Notice that the hermiticity of the interaction Hamiltonian Eq. (\ref{eq:interaction_ham}) together with the fermionic nature of the electronic operators ensure the property 
%For interactions that fulfill $V_{\bold{k}\bold{k}'} = V_{-\bold{k}-\bold{k}'}$, as it is the case for the Coulomb interaction considered here, one can further show that the gaps are odd under time-reversal, i.e. 
$\Delta_{\bar\tau \bar\sigma}(\bar{\bold{k}}) = -\Delta_{\tau \sigma}(\bold{k})$, i.e., that the gap functions are  odd under time reversal. 
% Therefore, the spin and valley index can be combined into the pseudo-spin index $i = \tau \cdot \sigma$ and the sums over $\tau$ and $\sigma$ can be reduced to a single sum over $i$. 
%\footnote{In this context we refer to spin $\uparrow$ as $+1$ and to spin $\downarrow$ as $-1$.}
%This pseudo-spin is invariant under time-reversal and the time-reversed product will be denoted as $\Bar{i}$. From now on we will also use the following convention:
%
%\begin{equation}
 %   i = \left\{ 
  %  \begin{matrix}
   % 1 & ,\ \text{for:} & i \equiv (K, \uparrow); & \bar{1} & ,\ \text{for:} & \bar{i} \equiv (K', \downarrow)\\
%    2 & ,\ \text{for:} & i \equiv (K, \downarrow); & \bar{2} & ,\ \text{for:} & \bar{i} \equiv (K', \uparrow)
 %   \end{matrix}
  %  \right. .
%\end{equation}
%
%The index $i=1,\bar{1}\ (2,\bar{2})$ can now also be associated to the upper (lower) band at the $K$ and $K'$ valley. With this we can finally set up the full mean-field Hamiltonian 
This property allows us to write the total grandcanonical mean field Hamiltonian in terms of the pseudospin index $(i,\bar{i})$ 
%$i=1,2$ and $\bar{i}=\bar{1}, \bar{2}$ 
introduced in the previous section 
\begin{align}
\label{eq:mean_field_final}
    \hat{H}^{\text{mf}}-\mu \hat{N}- C = &\sum\limits_{\bold{k},i=1,2}\xi_i(\bold{k}) \left(
	\hat{c}^\dagger_{\bold{k}i}\hat{c}^{}_{\bold{k}i}+\hat{c}^\dagger_{\bar{\bold{k}}\, \bar{i}}\hat{c}^{}_{\bar{\bold{k}}\,   			\bar{i}}
	\right) \notag \\
	-&\sum\limits_{\bold{k},i=1,2}\left( \Delta_i(\bold{k})\hat{c}^\dagger_{\bold{k}i}\hat{c}^\dagger_{\bar{\bold{k}}\,  				\bar{i}}+h.c.
	%\Delta_i^*(\bold{k}) \hat{c}^{}_{-\bold{k}\, \bar{i}}\hat{c}_{\bold{k} i}
	\right)\,,
\end{align}
where the dispersion relative to the chemical potential $\xi_i(\bold{k})$ was given in Eq. (\ref{eq:hyperbolic_dispersion_mu}).
It is noteworthy that for each  of the two pseudospins,   $(1,\bar{1})$ and $(2,\bar{2})$, the above expression has the BCS form, where the collective index $i$ plays  the role of the conventional spin. Thus, the mean field Hamiltonian can be readily diagonalized by a conventional   Bogoliubov transformation accounting for the full quasiparticle spectrum of the two-bands superconductor. 
\subsection{Bogoliubov transformation and coupled gap equations}
\label{sub:bologoliubov}
For a quadratic Hamiltonian like in Eq. (\ref{eq:mean_field_final}) the  Bogoliubov transformation has the form 
 \begin{equation}
    \begin{pmatrix}
    \hat{\gamma}^{}_{\bold{k}i}\\
    \hat{\gamma}_{\bar{\bold{k}}\, \bar{i}}^\dagger
    \end{pmatrix}
    = 
    \begin{pmatrix}
    u_{\bold{k}i} & -v_{\bold{k}i}\\
    v^*_{\bold{k}i} & u^*_{\bold{k}i}
    \end{pmatrix}
    \begin{pmatrix}
    \hat{c}^{}_{\bold{k}i}\\
    \hat{c}^\dagger_{\bar{\bold{k}}\, \bar{i}}
    \end{pmatrix},
    \ \
    |u_{\bold{k}i}|^2 + |v_{\bold{k}i}|^2 = 1\,,
    \label{eq:BogoliubovTrafo}
\end{equation}
where the condition on the sum of the coefficients  $v_{\bold{k}i}$ and  $u_{\bold{k}i}$ ensures the proper fermionic anticommutation relations of the quasi-particle operators $ \hat{\gamma}^{}_{\bold{k}i}$. By choosing for the coefficients  the conventional BCS form  
\begin{align}
 \label{eq:u_and_v}
   | u _{\bold{k}i}|^2= \frac{1}{2}\left( 1+\frac{\xi_i(\bold{k})}{E_i(\bold{k})}\right) , \quad | v_{\bold{k}i}|^2= \frac{1}{2}\left( 1-\frac{\xi_i(\bold{k})}{E_i(\bold{k})}\right)\,,
   \end{align}
   with $
    E_i(\bold{k}) = \sqrt{\big( \xi_i(\bold{k}) \big)^2 + \big| \Delta_i(\bold{k}) \big| ^2} $, 
we obtain the diagonalized mean-field Hamiltonian
\begin{align}
    \hat{H}^{\text{mf}} -\mu \hat{N} - C = \sum\limits_{\bold{k},i} E_i(\bold{k}) \left(
    \hat{\gamma}_{\bold{k}i}^\dagger \hat{\gamma}^{}_{\bold{k}i} + \hat{\gamma}_{\bar{\bold{k}}\, 
    \bar{i}}^\dagger \hat{\gamma}^{}_{\bar{\bold{k}}\, \bar{i}}
    \right)\,.
    \label{eq:QuasiparticleDispersion}
\end{align}
The simple and elegant expression above allows one to evaluate all the thermodynamic properties of the two-bands superconductor. 

Of primary interest for us are the self-consistent equations for the gaps $\Delta_1$ and $\Delta_2$. In particular, we are asking if nontrivial solutions exist, and if they are compatible with a realistic parametrization for NbSe$_2$ or for other Ising  superconducting TMDCs.  Bogoliubov operators describe excitations in the superconductor in terms of an ensemble of  non-interacting quasiparticles. The equilibrium occupation of the quasiparticle states is thus provided by the Fermi function according to 
$\langle \hat{\gamma}_{\bold{k}i}^\dagger \hat{\gamma}^{}_{\bold{k}i}\rangle =f(E_i(\bold{k}))$, and $\langle \hat{\gamma}_{\bold{k}i} \hat{\gamma}^{\dagger}_{\bold{k}i}\rangle =1-f(E_i(\bold{k}))$. 
 Starting from the definition of the gaps in  Eq. (\ref{eq:definition_gaps}) and expressing the averages in $\Delta_{\tau \sigma}^{\text{intra/inter}}(\bold{k})$ in terms of expectation values of Bogoliubov operators, yields  for the two-component vector 
 $\boldsymbol{\Delta}=(\Delta_1(\bold{k}), \Delta_2(\bold{k}))^T$ the coupled set of equations 
 
 %Doing so, one finally ends up with the coupled gap equation that assumes a very similar form as the usual gap equation obtained in conventional BCS theory for a single band. The self-consistent gap equation reads
%
\begin{align}
\label{eq:GapEquation}
   \boldsymbol{ \Delta}(\bold{k}) &= -\sum_{\bold{k}'}
   \cal{M}(\bold{k}') \boldsymbol{\Delta}(\bold{k}')\,,
   \\
  \cal{M}(\bold{k}') =&
    \begin{pmatrix}
    \intrakk \chi_1(\bold{k}')  && - \interkk \chi_2(\bold{k}') \\
    - \interkk \chi_1(\bold{k}') &&  \intrakk \chi_2(\bold{k}')
    \end{pmatrix}\,.
\end{align}
The functions $\chi_i(\bold{k}')$ incorporate the Fermi statistics of the quasiparticles and are given by
\begin{equation}
    \chi_i (\bold{k}) = \chi_i(\xi_i(\bold{k})) = \dfrac{1}{2E_i(\bold{k})} \tanh\left(\dfrac{\beta E_i(\bold{k})}{2}\right)\,,
    \label{eq:Chi}
\end{equation}
with $\beta=1/k_{\text{B}}T$, $k_\text{B}$ Boltzmann constant and $T$  the temperature. 
Once the solution for the gaps within the $K$ valley is found,  the gaps within the $K'$ valley follow from  $\Delta_{\bar{i}}(\bold{k})= -\Delta_i(-\bold{k})$.
In the following we shall refer to $\Delta_1$ and $\Delta_2$ as the gaps of the outer and inner Fermi surfaces, respectively, of the $K$ valley. 

%%%%%%%%%%%%%%%%%%%%%%%%%%%
\subsection{Temperature dependence of the inner and outer gaps}
\label{sub:T_dependence}
%%%%%%%%%%%%%%%%%%%%%%%%%%%%%%
%First discuss the analytical solutions at zero temperature and Tc. Then discuss the numerical results and show various parameter sets (dependence on  $\mu$, SOC, ratio Vintra/Vinter).
%Universality of the curves if Deltas and T divided by Tc?
Our first task will now be to find nontrivial solutions  of the gap equation  (\ref{eq:GapEquation}). For general $\bold{k}$-dependent interaction potentials this is a quite difficult task, as it is already the case for the conventional BCS gap equation. Here  the off-diagonal terms, introduced by a non-vanishing intervalley potential $\interkk$, give rise to a coupling between the $\Delta_1(\bold{k})$ and $\Delta_2(\bold{k})$ gaps which further complicates matters.
%\footnote{We will later see that for $\interkk \equiv 0$ the gap equation only allows trivial solutions, if $\intrakk>0$ as it is the case for the Coulomb repulsion.}
For this reason we shall focus on constant interactions in the following discussion and remember that $\intrakk, \interkk$ describe  long- and short-ranged parts, respectively, of the Coulomb repulsion in real-space. Qualitatively, these potentials can be conveniently described in terms of the screened interactions \cite{Roldan:prb2013}
\begin{align}
    \Uintrakk & := \intrakk \,N\Omega \approx \dfrac{2\pi e^2}{\epsilon (|\bold{k}-\bold{k}'| + q_{\text{TF}})} \,,\label{Vintra0}  \\
  \Uinterkk & :=   \interkk \,N\Omega = U_{\text{inter}}\, . \label{Vinter0}  
\end{align}
In Eq. (\ref{Vintra0})  $\epsilon$  is the dielectric constant of the environment, and $q_{\text{TF}} = 2\pi e^2 \rho(\mu)/\epsilon$ 
the Thomas-Fermi momentum which  describes the screening of  the long-range tail of the Coulomb interaction. 
In this case $\rho(\mu) = \rho_1(\mu)+\rho_2(\mu)$ is  the total DOS at the 
Fermi level, given by the sum of both DOS in Eq. (\ref{eq:DOS_i}), since we neglected contributions coming
from the $\Gamma$ pockets. 
In the second equality, Eq. (\ref{Vinter0}), the quantity  $U_{\text{inter}}$ is of 
the order of the product  $(e^2/2a)\Omega$, with $a$ the interatomic distance and $\Omega$ the size of the unit cell.   It  describes the leading term in the short-ranged Coulomb 
interaction of two electrons with opposite spin which occupy the same Nb $4d$ orbital. Although the exact value of $U_{\text{inter}}$ is not known,  it lies in the  few eV range \cite{Guinea:prb2012}.   \\
The expression 
$\Uintrakk$ can further be simplified by assuming that $q_{\text{TF}}$ is much larger than all of the considered 
exchanged momenta $|\bold{k}-\bold{k}'|\sim \bold{k}_{F,i}$ which are of the order of the Fermi momentum 
for one of the two bands. Note that this assumption can only be justified for not too large $\epsilon$ 
which would otherwise yield a rather small value for $q_{\text{TF}}$. Here we shall simply assume that this is 
indeed fulfilled. Hence, the intravalley potential also assumes a constant form, 
%\footnote{The only difference between the quantities $V$ and $U$ is the constant factor $A$, the size of the sample. Due to this we will not explicitly differentiate between these two anymore and will refer to both as potentials.}
%
\begin{align}
    \Uintrakk \approx \frac{2\pi e^2}{\epsilon q_{TF}} = \frac{1}{\rho(\mu)} \equiv U_{\text{intra}}\,.
\end{align}

Due to the now constant interactions and the fact that the right hand side of the gap equation in Eq. (\ref{eq:GapEquation}) only depends on $\bold{k}$ via the two potentials, the gap vector will be isotropic {\it{within each valley}}, i.e. $\bold{\Delta}(\bold{k}) =  \bold{\Delta}$.  Defining $\alpha_i := \frac{1}{N\Omega}\sum_{\bold{k}} \chi_i(\bold{k})$, the gap equation can now be written in the compact form
\begin{align}
    \mathbf{M}\bold{\Delta} =
    \left[ \mathbf{1} + 
		\begin{pmatrix}
			\ \, \Uintra \alpha_1 && -\Uinter \alpha_2 \\
			-\Uinter \alpha_1&& \ \,\Uintra \alpha_2
		\end{pmatrix}
	\right]
	 \bold{\Delta}
		= 0\,.
		\label{eq:compactGapEq}
\end{align}
Non-trivial solutions of the gap equation require a vanishing determinant of the matrix $\mathbf{M}$. By introducing the  potential   $\UattrI = (\Uintra^2-\Uinter^2)/\Uintra$, we obtain 
\begin{align}
    \det \mathbf{M} &= \UattrI\Uintra \alpha_1\alpha_2 + \Uintra(\alpha_1+\alpha_2)+1 = 0 \,.
    \label{eq:determinant}
\end{align}
 Since both  $\alpha_i$'s and the potentials $\Uintra,\Uinter$ are positive, the above condition can only be fulfilled if $\UattrI < 0$, i.e.  $\Uintra <  \Uinter$, leading to an attractive potential. Note that since the intervalley scattering enters quadratically in the definition of $U_\text{attr}$, its sign is not relevant here. Also, in the absence of intervalley scattering, $U_{\text{inter}}=0$, only the trivial solution $\boldsymbol{\Delta}=0$ exists for repulsive intravalley interaction.  The phase diagram for various combinations of the signs and relative strengths of the interactions is discussed in Sec. \ref{sub:symmetries}. 
 
 The sum over momenta  in the definition of the quantities $\alpha_i$ can be evaluated by transforming it into an energy integral    
\begin{align} 
    \alpha_i &= \dfrac{1}{N\Omega}\sum\limits_{\bold{k}} \chi_i(\bold{k}) = \int_{-\Lambda}^{\Lambda} \dif\xi\, \left(\rho_i(\mu) - d_i \xi \right)\chi_i(\xi)\theta(\xi_i^0-\Lambda) \notag\\
    &= \rho_i(\mu) \int_{-\Lambda}^{\Lambda} \dif\xi\, \chi_i(\xi)\theta(\xi_i^0-\Lambda) \,,
    %\chi_i(\xi) &= \dfrac{\tanh\left({\dfrac{\beta}{2}\sqrt{\xi^2+|\Delta_i|^2}}\right)}{2\sqrt{\xi^2+|\Delta_i|^2}},
\end{align}
%
%where \textcolor{blue}{$\vartheta(\xi_i^u)=\theta(\xi_i^u)\theta(\xi_i^u-\Lambda)$ is a product of Heaviside functions which replaces the $\theta$ function in the DOS from Eq. (\ref{eq:DOS_i}). 
where 
the Heaviside function ensures that the integration interval is below the top of the valence bands. Since we assume that $\Lambda$ defines a small energy interval around the Fermi energy, this requirement is automatically satisfied. 
%%ince the integration interval is rather small and we will only consider the cases where this integral is indeed symmetric, i.e. we are not to close to the top of the bands such that $\xi_i^u > \Lambda$.  
Usually in the computation of the integral one would approximate the DOS $\rho_i(\varepsilon)$ with its value $\rho_i(\mu)$ at the Fermi level. Here,  we naturally only get contributions from the latter since the linear term in the DOS leads to an odd integrand which vanishes upon integration.

Let us now for a moment consider NbSe$_2$ with a strongly shifted Fermi level, close to the top of the valence bands. According to the definition Eq. (\ref{eq:DOS_i}), the DOS at the Fermi level $\rho_i(\mu)$ vanishes if  $\xi_i^0<0$. We can thus differentiate between two configurations which could possibly lead to finite $\alpha_i$. In the first one,  the  chemical potential $\mu$  lies   between the top of the upper and the lower band ($\xi_1^0 > 0,\ \xi_2^0 < 0$); in the second one $\mu$  is below the top of both bands ($\xi_i^0> 0$).
%\footnote{Here, we neglected the influence of the energy cutoff $\Lambda$ in the second step function of $\vartheta(\xi_i^u)$ since it is usually small compared to the top of the band, i.e. $\xi_i^u-\Lambda \approx \xi_i^u$.}
In the first scenario  $\alpha_2 = 0$ and Eq. (\ref{eq:determinant}) leads to the familiar BCS gap equation, however now with the repulsive interaction $\Uintra>0$. As it is well-known, the BCS gap equation allows  non-trivial solutions only for attractive potentials. Hence in this  case the condition for finite gaps in Eq. (\ref{eq:determinant}) is not fulfilled and  Eq. (\ref{eq:compactGapEq}) is only solved by $\Delta_i = 0$. This means that a superconducting phase cannot exist for such range of chemical potentials.  Only the second scenario, where both interactions can take place, is relevant for the further discussion.
Notice that this interdependence of the two gaps implies that the there exists a single critical temperature $T_c$ at which both gaps vanish. Its expression is derived in analytical form below.
%Similar to the conventional BCS case, one can now obtain analytical results for the critical temperature and an equation whose solution yields the zero temperature gaps in the regime where $\vartheta(\xi_i^u)=1$, i.e.,  far below the top of both bands.
%
\subsubsection{The critical temperature}
At the critical temperature both gaps will have vanished and the superconducting state collapses. Hence, we simply set $\Delta_i = 0$ in $\alpha_i(T_c)$. With the assumption that the energy cutoff is much larger than the critical temperature, i.e. $\Lambda \gg k_B T_c$, the integrals can be solved analytically 
\begin{equation}
    \alpha_i (T_c)= \rho_i(\mu) \int_0^{2\Lambda/k_B T_c} \dif x\, \dfrac{\tanh x}{x} \approx -\rho_i(\mu) \ln{\dfrac{T_c}{\theta}},
\end{equation}
with the characteristic temperature $\theta = 2e^\gamma \Lambda/\pi k_B$ and the Euler-Mascheroni constant $\gamma \approx 0.577$. Inserting these expressions into Eq. (\ref{eq:determinant}), we find a polynomial of degree two in $\ln{\frac{T_c}{\theta}}$ whose solutions can be used to obtain the critical temperature\cite{noteTwoRoots}
\begin{align}
    \dfrac{T_c}{\theta} = 
    \exp\left[
			\dfrac{1}{2\UattrI}  \dfrac{1}{r}
			\left(
				1 + \sqrt{1- 2\dfrac{\UattrI}{\Uintra} \dfrac{r}{\bar{\rho}} }\
			\right)
		\right]\,.
\end{align}
Note that the DOS of the two bands $\rho_1(\mu)$ and $\rho_2(\mu)$ only enter in the symmetrical combinations 
\begin{equation}
    \dfrac{1}{r}= \frac{1}{ \rho_1} + \frac{1}{ \rho_2} \,, \quad \bar{\rho} = \dfrac{\rho_1 +\rho_2}{2}\,.
\end{equation}
%\textcolor{blue}{Estimate of $T_c$ ....} \\
When for similar DOS we approximate the value of $1/r$ by $2/\bar{\rho}$, the simplified expression for $T_c$ reads
\begin{equation}
\label{eq:Tc-simplified}
T_c \simeq \theta\;\exp\left(-\frac{1}{\bar{\rho}(\Uinter - \Uintra)}\right). 
\end{equation}
For repulsive interactions this result stresses again the importance of the local repulsion winning over the long-range one, encoded in our $\Uinter > \Uintra$ requirement. As we shall discuss in Sec.~\ref{sec:gap-symmetries}, depending on the signs and relative strengths of the potentials we may have to choose the other root of Eq.~\eqref{eq:determinant}. 
\\
As the temperature is lowered below $T_c$, the gaps start to grow and reach their maximal value at zero temperature. Some properties of the zero temperature gaps are elucidated below.
%%%%%%%%%%%%%%%%%%%%%%%%%%%%%%%%%%%%%%%%%%%%%%
\subsubsection{Zero temperature gaps}
%%%%%%%%%%%%%%%%%%%%%%%%%%%%%%%%%%%%%%%%%%%%%
The gaps at absolute zero cannot be given in an explicit closed form. One can, however, derive an equation whose solution yields them. Here, instead of writing $\Delta_i(T=0)$, we simply keep the notation $\Delta_i$ and remember that these are actually the gaps at $T=0$. Then, as the temperature approaches zero, the hyperbolic tangent reaches one and the integral  yielding $\alpha_i(T=0)$ can again be computed analytically. Further assuming $\Lambda \gg |\Delta_i|$, one finds the asymptotic form 
\begin{align}
\label{eq:alphas-Deltas}
    \alpha_i = \rho_i \int_0^{\Lambda/|\Delta_i|} \dif x \dfrac{1}{\sqrt{1+x^2}} \approx -\rho_i \ln \dfrac{|\Delta_i|}{2\Lambda} := -\rho_i  x_i\,,
\end{align}
with the dimensionless quantity $x_i = \ln \frac{|\Delta_i|}{2\Lambda}$.
By making use of the condition in Eq. (\ref{eq:determinant}) one can first get to a relation between $|\Delta_1|$ and $|\Delta_2|$. Afterwards,  using the first or the second row of the gap equation (\ref{eq:compactGapEq}), leads to 
\begin{align}
    &\Uinter \exp \left[
			\dfrac{1}{\rho_{j} \Uintra} \dfrac{1-\rho_{i}\Uintra  x_i} {1-\rho_{i} \UattrI  x_i}
			\right] \notag \\
			&+ \Uintra (1-\rho_i \UattrI x_i) e^{x_i} = 0\,.
\end{align}
In the above equation the index $j$ always refers to the opposite of $i$, i.e. if $i=1\ (2)$ then $j=2\ (1)$. When solving numerically the above expression in terms of $x_i$ we can thus get to the absolute values of both gaps $\Delta_i$.

By considering only the leading terms in the gap equation \eqref{eq:compactGapEq} we can find closed analytical expressions for the average gap $\bar{\Delta} := (\Delta_1 + \Delta_2)/2$ and the   
gap difference $\delta\Delta:=(\Delta_1 - \Delta_2)/2$. Using the the DOS difference $\delta\rho:=(\rho_1-\rho_2)/2$ and the already introduced average DOS $\bar{\rho}=(\rho_1+\rho_2)/2$, we can now add and subtract the two rows of Eq.~\eqref{eq:compactGapEq}, arriving at 
\begin{align*}
\bar{\Delta} \simeq (\Uintra-\Uinter) \;(\bar{\rho} \bar{\Delta} + \delta\rho\, \delta\Delta)\,\ln\left\vert\frac{\bar{\Delta}}{2\Lambda}\right\vert, \\[2mm]
\delta\Delta \simeq (\Uintra+\Uinter) \;(\bar{\rho} \delta\Delta + \delta\rho\, \bar{\Delta})\,\ln\left\vert\frac{\bar{\Delta}}{2\Lambda}\right\vert ,
\end{align*}
where we have approximated $\ln(\Delta_i/(2\Lambda))\approx \ln(\bar{\Delta}/(2\Lambda))$.
Keeping only the leading $\bar{\rho}\bar{\Delta}$ term in the first equation, we find
\begin{equation}
\label{eq:averageGap}    
\left\vert\bar{\Delta}\right\vert \simeq 2\Lambda\;\exp\left(-\dfrac{1}{\bar{\rho}\,(\Uinter-\Uintra)} \right)    .
\end{equation}
The expression for the average gap $\bar{\Delta}$ is highly reminiscent of the standard BCS result, linking in the same way the zero temperature gap and the critical temperature in Eq.~\eqref{eq:Tc-simplified}.\\ 
From the second equation, when using the result of \eqref{eq:averageGap}, we obtain
\begin{equation}
\label{eq:gapDifference}
\delta\Delta \simeq -\,\bar{\Delta}\,\frac{\delta\rho}{\bar{\rho}}\; \frac{\Uinter+\Uintra}{2\,\Uintra}.
\end{equation}
The difference between the two gaps is proportional to the difference between their normal DOS, which is reasonable. A less intuitive property of $\delta\Delta$ is that its sign is opposite to that of $\delta\rho$  -- in consequence, the band with larger $\rho_i(\mu)$ develops a smaller gap. Mathematically this is caused by the negative value of $\ln|\bar{\Delta}/(2\Lambda)|$. Physically, lower DOS in band $i$ means that the intravalley scattering, where both initial and final states are few, has lower amplitude than the intervalley scattering, where the final states belong to the band $j$, with higher DOS. The contrary is true for the band $j$.\\
Note that both \eqref{eq:averageGap} and \eqref{eq:gapDifference} were derived assuming that $\bar{\Delta}>\delta\Delta$. For an $s$-wave pairing the two gaps have opposite signs (as will be discussed in Sec.~\ref{sec:gap-symmetries}) and we would have to repeat our calculation for $\delta\Delta>\bar{\Delta}$ instead.

%%%%%%%%%%%%%%%%%%%%%%%%%
\subsubsection{Numerical results and comparison with experiments}
%%%%%%%%%%%%%%%%%%%%%%%
\label{sec:gap-symmetries}

\begin{figure}[t]
	\begin{center}
		\includegraphics[width=\columnwidth]{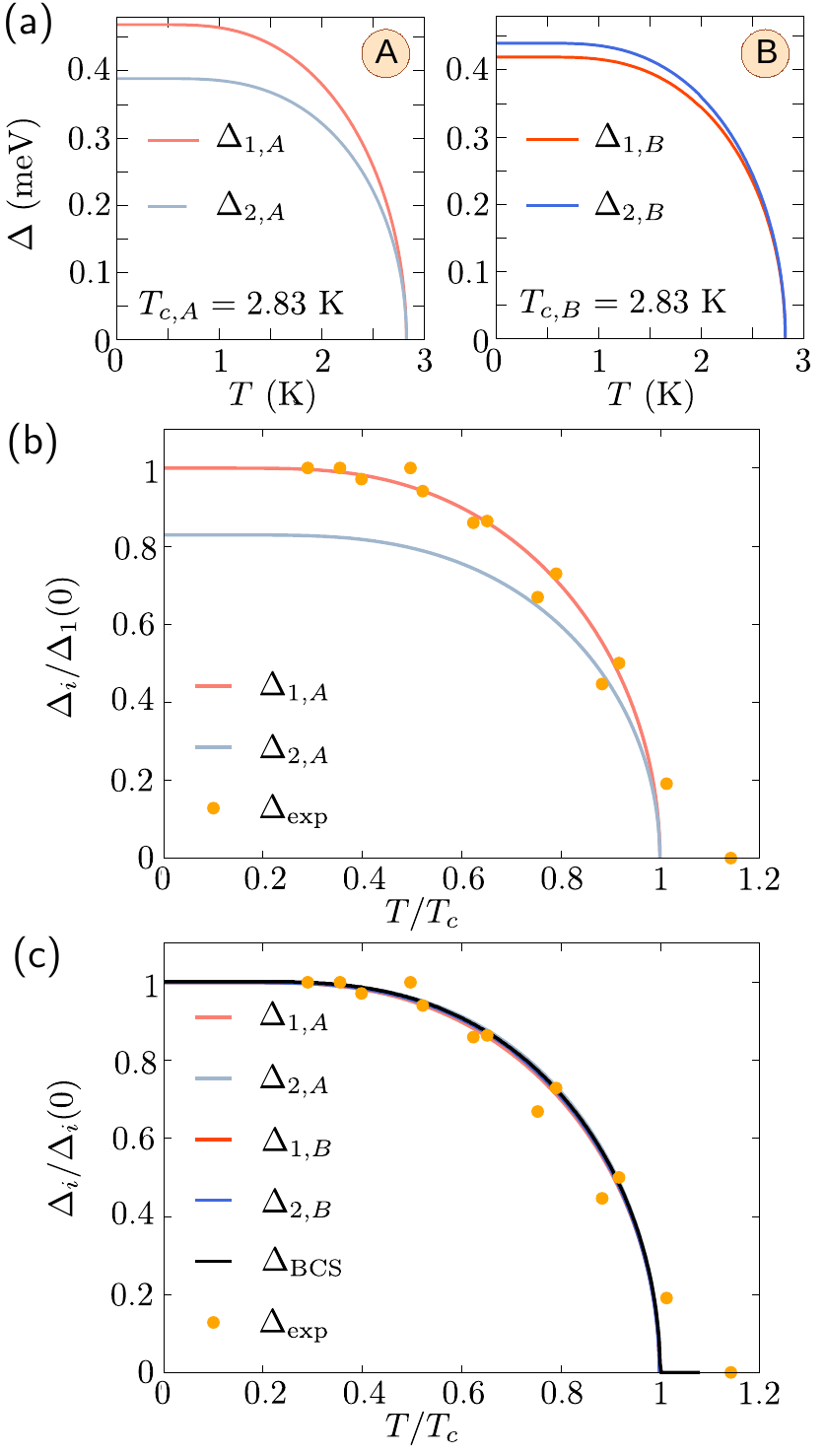}
		\caption{\small{Temperature dependence of the outer and inner  gaps, $\Delta_1$ and $\Delta_2$, respectively and universal scaling. (a) Numerical results showing the gaps' evolution towards a common critical temperature. The two distinct parametrizations  $A$ and $B$ in tables \ref{table:parameters} and \ref{table:sc-parameters-gaps} were used.  (b) Rescaled  theoretical results for the parametrization $A$ and comparison with  experimental points. The latter are digitized data from Fig.~3 in Ref.~[\onlinecite{Wan:arxiv2021}]. (c) A universal scaling of the gaps, independent of the chosen parametrization, is observed when dividing the theoretical gaps by their zero temperature value, and the temperature by the critical temperature.  }}
		\label{fig:numerics-vs-exp}
	\end{center}
\end{figure}

Getting to the full temperature dependence of the gaps requires numerical methods. Here, we used a self-consistent algorithm to solve the gap equation  (\ref{eq:GapEquation}) together with the determinant condition in Eq. (\ref{eq:determinant}).  The cutoff $\Lambda$ and the intervalley potential $\Uinter$ are free parameters. %They are  fixed in such a way that the resulting zero temperature gaps and the critical temperature are in the experimental range.  %Noat et al. \cite{Noat:prb2015} provide a comparison between several values for the gaps in the bulk as obtained by photoemission experiments. Although these values are not in perfect agreement with each other, they still lie around 1~meV. 
%
%Due to the missing experimental values for the gaps in monolayer NbSe$_2$, we simply assume that they drop by a similar factor as $T_C$, i.e. by a factor $\sim 2$. Hence, we infer that they can be expected around 0.5 meV.
 They are fixed by requiring the critical temperature to be $T_c\simeq 2.83$~K and the average gap at zero temperature  $\bar{\Delta}\simeq 0.4$~meV, in line with experimental estimates.
While the critical temperature is $T_c \approx 7.2$K for bulk NbSe$_2$,  it decreases with the number of layers \cite{Dvir:natcomms2018}. 
For example, Xi et al. \cite{Xi:natphys2016} give the critical temperature for both the bulk and the monolayer system which are $T_c^{\text{bulk}} \approx 7.0\,$K and $T_c^{\text{mono}} \approx 3.0\,$K for exfoliated NbSe$_2$ monolayers.
 For molecular beam epitaxy grown  monolayers the critical temperature has been found to vary between $T_c=0.9-2.4$~K \cite{Ugeda:natphys2016,Xing:nanolett2017,Zhao:natphys2019,Wan:arxiv2021}. 
A temperature dependence of the tunneling density of states, from which the average zero temperature gap was estimated to be around 0.4~meV is provided  in Ref.~[\onlinecite{Wan:arxiv2021}]. 
We notice that, having fixed $\Lambda$ and $\Uinter$, our predictions for the in-plane critical magnetic field discussed in the next section are parameter free. In Fig.~\ref{fig:numerics-vs-exp} we provide  numerical results for the evolution of the two gaps with temperature according to the TB parametrizations by Kim et al. \cite{Kim:prb2017} and He et al. \cite{He:commphys2018}, denoted by $A$ and $B$, respectively, in Table~\ref{table:parameters}. For both models we set $\Lambda=10$~meV. Further parameters and the values of $\Delta_i(T=0)$ are listed in Table~\ref{table:sc-parameters-gaps}. 

\begin{table}[h!]
\centering
\begin{tabular}{c | c c c c c c }
set & $\Uintra$ & $\Uinter$ & $\Delta_1(0)$ & $\Delta_2(0)$ & $\bar{\Delta}^{(30)}$ & $\delta\Delta^{(31)}$\\ 
 \hline\hline
 $A$ & 11.83 & 18.04 & 0.469 & 0.388 & 0.442 & 0.0495 \\ 
 $B$ & 16.29 & 24.77 & 0.42  & 0.44 & 0.43 & -0.0123\\
\hline\hline
\end{tabular} 
\caption{
\label{table:sc-parameters-gaps}
Interaction parameters and resulting zero temperature gaps. The difference in the interaction strengths arises due to the different  values of the DOS at the Fermi level for the $A$ and $B$ sets. The units for the potentials are eV{\AA}$^2$, the units for the gaps are meV. The last two columns show the value of $\bar{\Delta}$ evaluated with Eq.~\eqref{eq:averageGap} and $\delta\Delta$ evaluated with Eq.~\ref{eq:gapDifference}.} 
\end{table}

%This further  
%gaps at different $\mu$, with critical chem. pot.
As shown in Fig.  \ref{fig:numerics-vs-exp}(a), we find two finite gaps with the same critical temperature and which assume their largest values at zero temperature. The size of the gaps slightly depends on the chosen parametrization but the qualitative behavior is rather similar. 
Fig. \ref{fig:numerics-vs-exp}(b) shows a comparison of the experimental data\cite{Wan:arxiv2021}, and the theoretical   predictions with the $A$  parametrization. Notice that the evolution of the data points is compatible with the existence of the two gaps. 
Finally, Fig. \ref{fig:numerics-vs-exp}(c) demonstrates  that our predictions become BCS-like and independent of the specific parametrization when the  temperature is rescaled by the critical temperature  and the gaps by their respective zero temperature values.\\
Our approximate formulae \eqref{eq:averageGap} and \eqref{eq:gapDifference} work well in both models. In the model $A$ the average gap $\bar{\Delta}$  differs by about 3\% from the numerical result, while in the model $B$ (where the two bands are more similar) the two results agree up to 1\%. The gap differences $\delta\Delta$ in both models are overestimated by about 20\%.\\
%
%\textcolor{blue}{Different SOC parameter (?)
%Different ratio of Intra, inter (?) }
%%
%\begin{figure}[ht]%
%	\begin{center}
%		\includegraphics[width=8cm,angle=0]{figures/different_chemPot.png}
%		\caption{\small{Placeholder:Temperature dependence of  the inner and outer gaps for different values of the Ising SOC strength. }}
%		\label{fig:T-dependence}
%	\end{center}
%\end{figure}
%

\begin{figure}[ht]
	\begin{center}
		\includegraphics[width=\columnwidth]{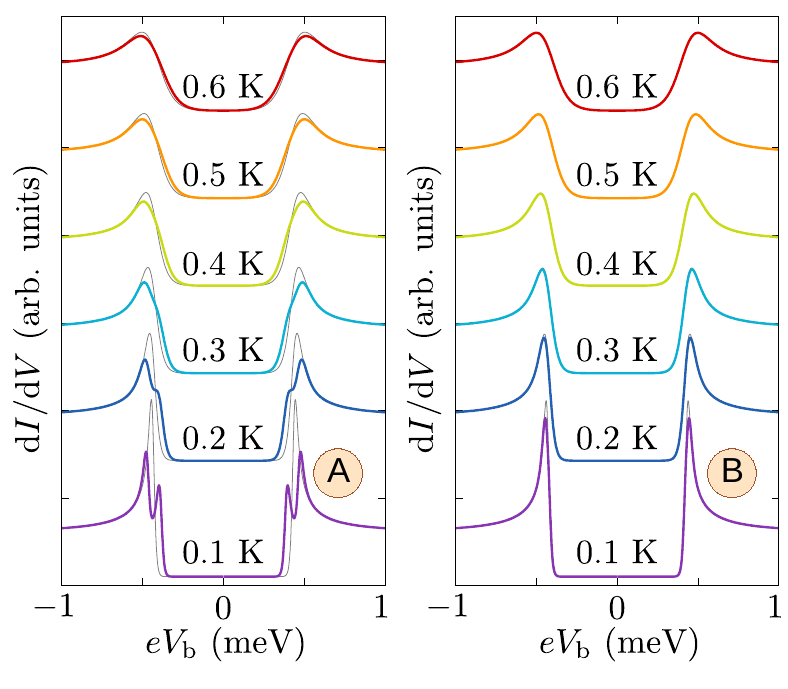}
		\caption{\small{Differential conductance vs bias voltage  of monolayer NbSe$_2$ at several temperatures. This quantity is proportional to the tunneling density of states.   In both tight-binding parametrizations $T_c = 2.83$~K. Thin grey lines correspond to the tunneling DOS with only one gap $\bar{\Delta}$. The two gaps are still recognizable at $T \approx 0.1\,T_c$ for the $A$, but not for the $B$ set. At higher temperatures, the gaps merge and are not visible in the differential conductance, which is also almost indistinguishable from the case of one gap $\bar{\Delta}$.
		%The points at $T=0.4$~K in the \textcolor{blue}{last panel} are digitized data from Fig.~3 in Ref. [\onlinecite{Wan:arxiv2021}]. Notice that in the experiment $T_c=2.0$~K.
		}}
		\label{fig:tdos}
	\end{center}
\end{figure}

It is by now well established that bulk NbSe$_2$ has two gaps, the second gap being due to the electrons in the Se pocket around the $\Gamma$ point \cite{Noat:prb2015}. Recent tunnel junction spectroscopic measurements of NbSe$_2$ devices with few layers have shown that the second gap grows weaker with decreasing number of layers,\cite{Dvir:natcomms2018} becoming invisible in a bilayer device\cite{Kuzmanovic:arxiv2021}. Likewise, the scanning tunneling microscope (STM)  measurements of  monolayer devices in Ref.~[\onlinecite{Wan:arxiv2021}] do not display clear signatures of a second gap. \\
In order to establish whether the second gap would be at all visible in the $\text{d}I/\text{d}V$ characteristics of an STM, we calculated the tunneling density of states\cite{Tinkham:book} for both parametrizations, $A$ and $B$, of our effective model. The results are  shown in Fig. \ref{fig:tdos}. 
The two gaps are clearly visible in the $A$ model at $T=0.1$~K. Nevertheless, as the temperature rises, their quasiparticle peaks merge and at $T=0.4$~K 
%the lowest temperature shown in Ref. [\onlinecite{Wan:arxiv2021}], 
only a shallower slope indicates that two gaps are present. With the $B$ parametrization the two gaps are too similar in magnitude to be clearly distinguishable even at $T=0.1$~K. Thus in order to distinguish the two gaps the tunneling experiments would have to be carried out at very low temperatures.
\subsection{Symmetries of the inner and outer gaps  }
\label{sub:symmetries}
\begin{figure}[ht]
	\begin{center}
		\includegraphics[width=\columnwidth]{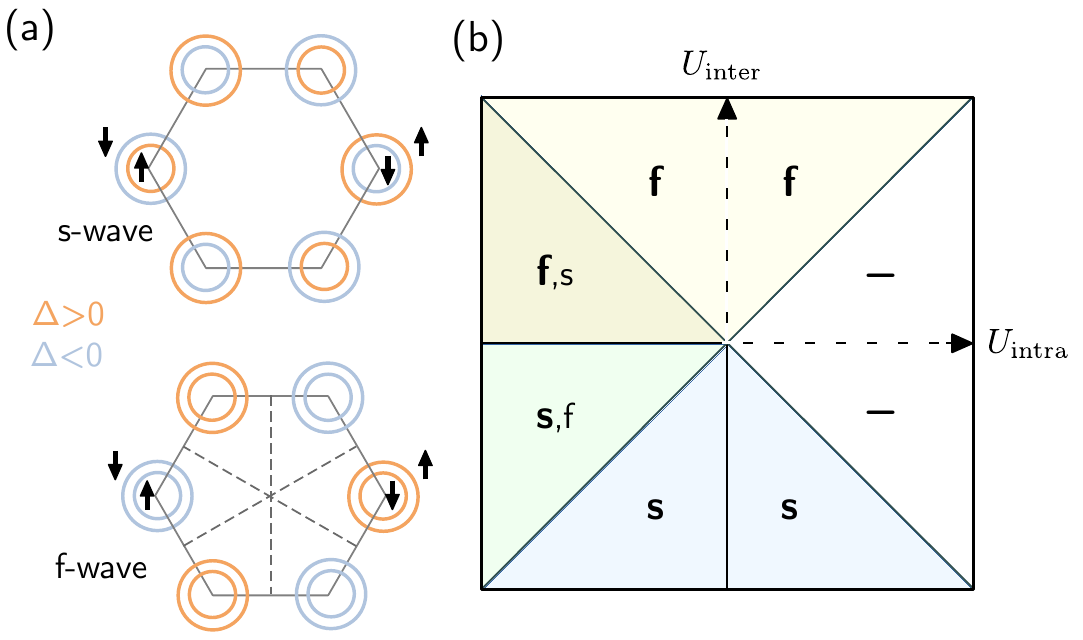}
		\caption{ Symmetries of the superconducting gaps induced by  spin conserving interactions in the absence of  spin-orbit coupling (the two Fermi surfaces in each pocket are drawn as split only for clarity). \small{(a) In a  hexagonal lattice with Fermi surfaces around the Dirac valleys the two  symmetries consistent with locally   isotropic gaps are of the $s$ or $f$ type. (b) Leading gap symmetries when two competing scattering processes can take place, depending on the sign and relative strength of $U^{\text{intra}}$ and $U^{\text{inter}}$,} and without SOC. When only one, repulsive, interaction is present, pairing does not occur.}
		\label{fig:gap-symmetries}
	\end{center}
\end{figure}

The symmetry of the gap is defined by its behavior under the reflection of its momentum, i.e. for $s,d$-wave symmetry $\Delta_{\tau\sigma}(\bold{k}) = \Delta_{\bar\tau\sigma}(-\bold{k})$ while for $p,f$-wave $\Delta_{\tau\sigma}(\bold{k}) = -\Delta_{\bar\tau\sigma}(-\bold{k})$. In systems with hexagonal lattice and small Fermi surfaces two types of symmetry allow an isotropic gap, the $s$ and $f$ symmetry\cite{Schaffer:jpcm2014}, as illustrated in Fig.~\ref{fig:gap-symmetries}(a). 
To see which one applies to our case, we recall that the fermionic anticommutation rules relate the gaps of opposite momenta and spins according to 
$
\Delta_{\tau\sigma}(\bold{k}) = -\Delta_{\bar\tau \bar\sigma}(-{\bold{k}})$. 
From the local isotropy of the gaps, it also follows  that 
\begin{equation}
 \Delta_{\tau\sigma}(\bold{k}) =    \Delta_{\tau\sigma}(-{\bold{k}})\,.
\end{equation}

Let us start by considering the easier case of zero SOC. 
In this case  $\chi_1(\bold{k}) = \chi_2(\bold{k})$ (which already implies $|\Delta_1|=|\Delta_2|$), and  hence  $\alpha:=\alpha_1=\alpha_2$ in Eq.~\eqref{eq:compactGapEq} can be factorized from the matrix, 
\begin{equation}
\left(\begin{array}{c} \Delta_1 \\[2mm] \Delta_2 \end{array}
\right)
= -\alpha
\left(\begin{array}{cc}
\Uintra & -\Uinter  \\[2mm]
-\Uinter & \Uintra
\end{array}
\right)
\left(\begin{array}{c} \Delta_1 \\[2mm] \Delta_2 \end{array}
\right).    
\end{equation}
The first row of this matrix equation yields
\begin{equation}
    \Delta_1 = \frac{1 + \alpha \Uintra}{\alpha \Uinter} \Delta_2,    
\end{equation}
while the requirement of the existence of non-trivial solutions, Eq. \eqref{eq:determinant}, implies
\begin{equation}
     \alpha_\pm = - \frac{1}{\Uintra \mp \Uinter}.
\end{equation}
Hence, $\Delta_1=\pm\Delta_2$; i.e.  in the absence of SOC the two gaps have the same amplitude but not necessarily the same sign. 
Since $\alpha$ is a sum of non-negative numbers, it must be positive. Depending on the sign and relative strength of $U^{\text{intra}}$ and $U^{\text{inter}}$ either one, none, or both of $\alpha_\pm$ are positive. Hence, according to the properties of the $s$ and $f$ gaps upon reflection of $\bold{k}$, we conclude  
\begin{equation}
\label{eq:gap-symmetries}
\begin{array}{l l l}
\alpha_+ > 0 & \quad\rightarrow\quad \Delta_1\,\Delta_2 > 0,
& \quad\textnormal{$f$-wave}, \\[2mm]
\alpha_- > 0 & \quad\rightarrow\quad \Delta_1\,\Delta_2 < 0, & \quad\textnormal{$s$-wave}.
\end{array}
\end{equation}
When both $\alpha_\pm>0$, the dominant  is that one  which results in greater energy gain upon condensation, i.e. in the larger amplitude of the gap. From Eq.~\eqref{eq:alphas-Deltas} we see that the smaller $\alpha$ results in the larger gap. Therefore $\alpha_+ \lessgtr \alpha_-$ results in a dominant gap with $f$ ($s$) symmetry.

When we include the effects of SOC, the two gaps become mixtures of the $s$ and $f$ character. In particular, because $\Delta_i(\bold{k})=-\Delta_{\bar{i}}(\bar{\bold{k}})$, then it holds for the average and difference gaps  
\begin{equation}
\label{eq:gap-symmetries-SOI}
\begin{array}{l l}
{\Delta}_K = \frac{1}{2}(\Delta_1+\Delta_2): &  \quad{\Delta}_K(\bold{k}) = -{\Delta}_{\bar{K}}(\bar{\bold{k}})\quad(f)\,, \\[2mm]
\delta\Delta_K = \frac{1}{2}(\Delta_1-\Delta_2): & 
\quad\delta\Delta_K(\bold{k}) = \delta\Delta_{\bar{\text{K}}}(\bar{\bold{k}})\quad(s)\,.
\end{array}
\end{equation}
Whether $\Delta_{K}$ or $\delta\Delta_K$ determines the prevalent symmetry depends on the sign of $\Delta_1\Delta_2$. In the limit of vanishing SOC we see again that with $\Delta_1\Delta_2>0$ the dominant gap is $\Delta_K$ i.e. $f$-wave, while for $\Delta_1\Delta_2<0$ the main gap is $\delta\Delta_K$ with $s$-wave symmetry. 
%
%%%%%%%%%%%%%%%%%%%%%%%%%%%%%%
\section{Effects of an in-plane field}
\label{Sec:magnetic-field}
%%%%%%%%%%%%%%%%%%%%%%%%%%%%%
We now turn to the effect of an in-plane field on the superconducting state of monolayer NbSe$_2$. We are mostly interested in the temperature dependence of the critical magnetic field $B_c(T)$. We start by first considering the system without electronic interactions but  with an additional magnetic field $\bold{B} = B \hat{e}_x$ along the $x$-axis. The single particle grandcanonical Hamiltonian at finite magnetic fields is 
\begin{align}
    \hat{H}_{\text{B}} -\mu \hat{N}= \sum\limits_{\bold{k}\tau\sigma} \xi_{\tau\sigma}(\bold{k}) \hat{c}_{\bold{k}\tau\sigma}^\dagger
                \hat{c}_{\bold{k}\tau\sigma}
                +\mu_{\text{B}} B \sum\limits_{\bold{k}\tau\sigma} \hat{c}_{\bold{k}\tau\sigma}^\dagger
                \hat{c}_{\bold{k}\tau\bar\sigma} \,,
\end{align}
where $\mu_{\text{B}}$ is the Bohr magneton. 
%and the dispersion $\xi_{\tau\sigma}(\bold{k})$ is now denoted with the spin and valley index instead of the pseudo-spin $i$.
The above Hamiltonian can be diagonalized using the Ansatz 
\begin{align}
    \begin{pmatrix}
        \hat{c}_{\bold{k}\tau\uparrow}\\
        \hat{c}_{\bold{k}\tau\downarrow}
    \end{pmatrix}
    =
    \begin{pmatrix}
        a_{\bold{k}\tau} & b_{\bold{k}\tau} \\
       -b_{\bold{k}\tau} & a_{\bold{k}\tau}
    \end{pmatrix}
    \begin{pmatrix}
        \hat{f}_{\bold{k}\tau +}\\
        \hat{f}_{\bold{k}\tau -}
    \end{pmatrix},\ \
    |a_{\bold{k}\tau}|^2+|b_{\bold{k}\tau}|^2 = 1\,.
    \label{eq:AnsatzDiagonalizationBField}
\end{align}
A possible choice  of the field-dependent parameters is
\begin{align}
    \left.
    \begin{matrix}
     a_{\bold{k}\tau} \\
     b_{\bold{k}\tau}
    \end{matrix}
    \right\} = 
    \pm \sqrt{ \dfrac{1}{2} \left( 1 \pm \dfrac{\delta\xi_\tau(\bold{k})}{\sqrt{(\delta\xi_\tau(\bold{k}))^2 + (\mu_B B)^2}}\right) }\,,
\end{align}
with  $2\delta\xi_\tau(\bold{k}) = \xi_{\tau\uparrow}(\bold{k})-\xi_{\tau\downarrow}(\bold{k})$.  It  results in the  eigenenergies
\begin{align}
    \DE{\tau n}(\bold{k}) &= \bar{\xi}_\tau(\bold{k}) \pm \sqrt{(\delta\xi_{\tau}(\bold{k}))^2+(\mu_B B)^2}\,,
    \label{eq:BfieldEigenenergies}
\end{align}
where
$2\bar{\xi}_\tau(\bold{k}) = \xi_{\tau\uparrow}(\bold{k})+\xi_{\tau\downarrow}(\bold{k})$. 
These energies depend on the index $n=+,-$ and not on the spin $\sigma$ anymore since the magnetic field is oriented perpendicularly to the spin quantization axis set by the Ising SOC. Remembering the time-reversal relation $\xi_{\bar\tau \bar\sigma}(\bar{\bold{k}}) = \xi_{\tau\sigma}(\bold{k})$ for the single-particle energies at zero magnetic field, one finds for the coefficients and energies the relations 
\begin{equation}
\label{eq:a_b}
    a_{\bold{k}\tau} = -b_{\bar{\bold{k}} \bar\tau}\,,\quad \DE{\tau n}(\bold{k}) = \DE{\bar\tau n}(\bar{\bold{k}})\,.
\end{equation}
In order to describe the influence of the magnetic field on the superconducting phase one now has to express the mean-field interaction term in Eq. (\ref{eq:MFInteractionTerm}) in the new eigenbasis, i.e. in terms of the operators $\hat{f}_{\bold{k}\tau n}$. Doing so, one eventually arrives at the full mean-field Hamiltonian $\hat{H} = \hat{H}_B + \hat{H}_{\text{int}}^{\text{mf}}$ describing superconductivity in an Ising spin-orbit coupled TMDC monolayer in the presence of an in-plane magnetic field, 
%\footnote{Since it is rather straight forward, we do not show the full derivation here. After going from the $\hat{c}$ to the $\hat{f}$ basis, one can again use the time-reversal relation for the gaps $\Delta_{\tau,\sigma}(\bold{k})$ and the aforementioned relation for the coefficients $a_{\bold{k}\tau}, b_{\bold{k}\tau}$ to compactify the result. By further shifting the summation from $\bold{k}$ to $-\bold{k}$ for the summands corresponding to the $K'$ valley ($\tau=-1$), one obtains the given expression.}
%
\begin{align}
    &\hat{H}-\mu \hat{N} -C = \sum\limits_{\bold{k},n} \DE{Kn}(\bold{k}) \left( \hat{f}^\dagger_{\bold{k}Kn}\hat{f}_{\bold{k}Kn}    
    +\hat{f}^\dagger_{\bar{\bold{k}}\bar{K}n}\hat{f}_{\bar{\bold{k}}\bar{K}n} \right)      \notag\\
    &+ \sum\limits_{\bold{k},n} \Delta_{Kn}(\bold{k}) \hat{f}^\dagger_{\bold{k}Kn}\hat{f}^\dagger_{\bar{\bold{k}}\bar{K}n}    + h.c.    \label{eq:Hamiltonian_SCAndB}  \\
    &+ \sum\limits_{\bold{k}} \Delta_{K\pm}(\bold{k}) \left( \hat{f}^\dagger_{\bold{k}K+}\hat{f}^\dagger_{\bar{\bold{k}}\bar{K}-} -
    \hat{f}^\dagger_{\bold{k}K-}\hat{f}^\dagger_{\bar{\bold{k}}\bar{K}+} \right) \notag
    + h.c. \,. 
\end{align}
Here, we have defined the three new effective gaps which can be expressed in terms of the gaps $\Delta_{\tau\sigma}(\bold{k})$ (they obey $\Delta_{K\uparrow}=-\Delta_{K'\downarrow}=\Delta_1$, $\Delta_{K\downarrow}=-\Delta_{K'\uparrow}=\Delta_2$). The three new gaps are 
\begin{align}
    \Delta_{K+}(\bold{k}) &= -(a_{\bold{k}K}^2 \Delta_{K\uparrow}(\bold{k}) + b_{\bold{k}K}^2 \Delta_{K\downarrow}(\bold{k})) \,,\\
    \Delta_{K-}(\bold{k}) &= b_{\bold{k}K}^2 \Delta_{K\uparrow}(\bold{k}) + a_{\bold{k}K}^2 \Delta_{K\downarrow}(\bold{k}) \,,\\
    \Delta_{K\pm}(\bold{k}) &= a_{\bold{k}K}b_{\bold{k}K} (\Delta_{K\uparrow}(\bold{k}) - \Delta_{K\downarrow}(\bold{k}))\,.\label{eq:effGap12}
\end{align}
The first two gap functions $\Delta_{Kn}(\bold{k})$ couple electrons of different $K$ valleys but the
same energy, while $\Delta_{K\pm}(\bold{k})$ describes a pairing of electrons with different energies. In the latter case, it depends on the amplitude of the magnetic field values and of  the superconducting pairing energies to which extent this term can contribute. Notice that Eq. (\ref{eq:a_b}) ensures that 
the new gaps $\Delta_{Kn}(\bold{k})$ have $f$-wave character while 
$\Delta_{K\pm}(\bold{k})$ is $s$-wave like, cf. also Eq. (\ref{eq:gap-symmetries-SOI}).  
\subsection{Diagonalization of the  mean field Hamiltonian in planar magnetic field}
To get to the gap equation which now includes the magnetic field we need to evaluate the averages in the definition of the order parameter in Eq. (\ref{eq:definition_gaps}). This requires to find the new set of Bogoliubov quasiparticles which diagonalize the Hamiltonian in Eq. (\ref{eq:Hamiltonian_SCAndB}). To this aim  we first rewrite it in the Bogoliubov-de Gennes (BdG) form 
\begin{align}
    \hat{H} - \mu \hat{N} - \tilde{C} = \sum\limits_{\bold{k}} \hat{\Psi}_{\bold{k}}^\dagger \hat{\mathcal{H}}_{\text{BdG}}
    (\bold{k})\hat{\Psi}_{\bold{k}}\,,
\end{align}
where we introduced the BdG Hamiltonian $\hat{\mathcal{H}}_{\text{BdG}}(\bold{k})$ and the Nambu spinor $\hat{\Psi}_{\bold{k}}$, respectively. They are given by 
\begin{align}
\hat{\mathcal{H}}_{\text{BdG}}(\bold{k}) &= 
\begin{pmatrix}
\DE{K+}(\bold{k}) & 0 & \Delta_{+}(\bold{k}) & \Delta_{\pm}(\bold{k}) \\ 
0 & \DE{K-}(\bold{k}) & -\Delta_{\pm}(\bold{k}) & \Delta_{-}(\bold{k}) \\
\Delta_{+}^*(\bold{k}) & -\Delta^*_{\pm}(\bold{k}) & -\DE{K+}(\bold{k}) & 0 \\
\Delta^*_{\pm}(\bold{k}) & \Delta^*_{-}(\bold{k}) & 0 & -\DE{K-}(\bold{k})
\end{pmatrix} \,, \label{eq:BdGMatrix}
\end{align}
where we used the abbreviation $\Delta_{Kn}=\Delta_n$, $\Delta_{K\pm}=\Delta_\pm$, and 
\begin{align}
\hat{\Psi}_{\bold{k}} &= 
\begin{pmatrix}
\hat{f}_{\bold{k}K+}, & \hat{f}_{\bold{k}K-}, & \hat{f}^\dagger_{-\bold{k}K'+}, & \hat{f}^\dagger_{-\bold{k}K'-}
\end{pmatrix}^T \,.
\label{eq:NambuSpinor}
\end{align} 
Getting the eigenvalues of the BdG matrix above is a  simple task. In contrast, finding the unitary transformation matrix $U$, i.e. the corresponding normalized eigenvectors, is  rather intricate.   Our way to get to their  analytic form is discussed  in Appendix A. In the following we are going to refer to the positive eigenenergies as $\DxiT{n}(\bold{k})$ (cf. Eq. (\ref{eq:EigenvaluesFiniteB2})), and to the entries of $U$ as $\DU{ij}$ (cf. Eq. (\ref{eq:unitaryTrafo})). The spinor %
$\hat{\Gamma}_{\bold{k}} = (
\hat{\gamma}_{\bold{k}K+},  \hat{\gamma}_{\bold{k}K-},  \hat{\gamma}^\dagger_{-\bold{k}K'+},  \hat{\gamma}^\dagger_{-\bold{k}K'-})^T
$, which contains the new Bogoliubov quasiparticle operators $\hat{\gamma}_{\bold{k}\tau n}$, is  related to the $\hat{f}$ operators by   $\hat{\Psi}_{\bold{k}} = U\hat{\Gamma}_{\bold{k}}$.  The Hamiltonian finally becomes diagonal in this basis
\begin{align}
    \hat{H} - \mu \hat{N} - \tilde{C} = \sum\limits_{\bold{k},n} \DxiT{n}(\bold{k}) \left( 
    \hat{\gamma}_{\bold{k}\tau n}^\dagger \hat{\gamma}^{}_{\bold{k}\tau n} + \hat{\gamma}_{\bar{\bold{k}}\, 
    \bar\tau n}^\dagger \hat{\gamma}^{}_{\bar{\bold{k}}\, \bar\tau n} \right)\,.
\end{align}
Notice that, according to Eq. (\ref{eq:EigenvaluesFiniteB2}), the quasiparticle spectrum now displays a quite intricate dependence on the three gaps $\Delta_n(\bold{k})$ and $\Delta_{\pm}(\bold{k})$. 
This suggests a multitude of different superconducting phases, possibly even of non-trivial topological character, in line with Ref.~[\onlinecite{Shaffer:prb2020}]. 
We postpone such analysis to a future work. 

The focus here is rather on the benchmark of the theory against available  data at finite magnetic field; explicitly, on the dependence of the critical magnetic field on temperature. As discussed above, having fixed the values of $\Lambda$ and $\Uintra$ to evaluate the temperature dependence of the zero-field gaps,  the theory is parameter free if we take a $g$-factor of $2$. Thus, an agreement with the experimental data will give us confidence in the   predictive power of the theory for future investigations.

%%%%%%%%%%%%%%%%%%%%%%%%%%%%%%%
\subsection{Gap equation for the critical in-plane field}
%%%%%%%%%%%%%%%%%%%%%%%%%%%%%%%%%%%%
To get the new gap equation we  express the operators $\hat{c}$, entering   the definition of the gaps $\Delta_{\tau\sigma}$, in terms of the new quasiparticle operators $\hat{\gamma}$, 
\begin{align}
    \begin{pmatrix}
        \hat{c}_{\bold{k}K\uparrow} \\
        \hat{c}_{\bold{k}K\downarrow} \\
        \hat{c}^\dagger_{-\bold{k}K'\uparrow} \\
        \hat{c}^\dagger_{-\bold{k}K'\downarrow}
    \end{pmatrix}
    &=
    \bold{V}(\bold{k})
    \begin{pmatrix}
        \hat{\gamma}_{\bold{k}K+} \\
        \hat{\gamma}_{\bold{k}K-} \\
        \hat{\gamma}^\dagger_{-\bold{k}K'+} \\
        \hat{\gamma}^\dagger_{-\bold{k}K'-}
    \end{pmatrix} \,.
    \label{eq:CGammaRelation}
\end{align}
The unitary transformation is defined by
\begin{align}
    \bold{V}(\bold{k}) &= 
    \begin{pmatrix}
       \DM_{\bold{k}K} & \bold{0}_{2\times2} \\
       \bold{0}_{2\times2} & \DM^*_{-\bold{k}K'}
    \end{pmatrix}
    U \,,
    \end{align}
with  $\DM_{\bold{k}\tau}$ the block matrix in Eq. (\ref{eq:AnsatzDiagonalizationBField}). It has elements 
    \begin{align}
   \bold{V}(\bold{k}) &=   \begin{pmatrix}
        \DV{11}(\bold{k}) & \DV{12}(\bold{k}) & \DV{41}^*(\bold{k}) & -\DV{42}^*(\bold{k}) \\
        \DV{21}(\bold{k}) & \DV{22}(\bold{k}) & \DV{31}^*(\bold{k}) & -\DV{32}^*(\bold{k}) \\
        \DV{31}(\bold{k}) & \DV{32}(\bold{k}) & -\DV{21}^*(\bold{k}) & \DV{22}^*(\bold{k}) \\
        \DV{41}(\bold{k}) & \DV{42}(\bold{k}) & -\DV{11}^*(\bold{k}) & \DV{12}^*(\bold{k}) 
    \end{pmatrix} \,.
    \label{eq:VTrafo}
\end{align}
By inserting the relations from Eq. (\ref{eq:CGammaRelation}) into the definition of the gaps in Eq. (\ref{eq:definition_gaps}), one can derive the new set of coupled gap equations for TMDC monolayers in an in-plane magnetic field. 
%\footnote{Here, we left away the valley index $\tau$ since we are only considering the gaps $\Delta_{K\sigma}$ in this case. The gaps in the $K'$ valley are then again given by the time-reversal relation mentioned earlier.}
%
They have the form in Eq. (\ref{eq:GapEquation}), with a magnetic field dependent matrix 
\begin{align}
\label{eq:GapEquation2}
   \boldsymbol{ \Delta}(\bold{k}) &= -\sum_{\bold{k}'}
  \tilde{\cal{M}}(\bold{k}',B) \boldsymbol{\Delta}(\bold{k}')\,.
\end{align}
Explicitly it holds 
\begin{widetext}
\begin{equation}
     \begin{pmatrix}
    \Delta_{1}(\bold{k}) \\
    \Delta_{2}(\bold{k})
    \end{pmatrix}
    = %\notag \\
    -\sum_{\bold{k}'}
    \left[
    \begin{pmatrix}
    \left( \intrakk \Dg{1}(\bold{k'}) - \interkk \Dh{1}(\bold{k'}) \right) \tilde\chi_1(\bold{k}') &
    \left( \intrakk \Dh{2}(\bold{k'}) - \interkk \Dg{2}(\bold{k'}) \right) \tilde\chi_2(\bold{k}') \\
    \left( \intrakk \Dh{1}(\bold{k'}) - \interkk \Dg{1}(\bold{k'}) \right) \tilde\chi_1(\bold{k}') &
    \left( \intrakk \Dg{2}(\bold{k'}) - \interkk \Dh{2}(\bold{k'}) \right) \tilde\chi_2(\bold{k}') \\
    \end{pmatrix}
    \begin{pmatrix}
    \Delta_{1}(\bold{k}') \\
    \Delta_{2}(\bold{k}')
    \end{pmatrix}
    \right],
    \label{eq:GapEqFiniteB}
\end{equation}
where the functions $\tilde\chi_1(\bold{k}')$, $\tilde\chi_2(\bold{k}')$ are the defined as in Eq. (\ref{eq:Chi}), but now with the new eigenenergies $\DxiT{+}(B)$, and 
$\DxiT{-}(B)$, respectively. 
Due to the action of the in-plane magnetic field, the elements of ${ \tilde{\cal{M}}}(\bold{k}',B)$   are a mixture of the original interactions ${V}^{\text{intra/inter}}_{\bold{k} \bold{k}'}$.  The energies  read, 
\begin{align}
    \DxiT{n} &=  \dfrac{1}{\sqrt{2}}\sqrt{\DET{+}^2+\DET{-}^2+2|\Delta_{\pm}|^2 \pm 
                \sqrt{ \left( \DET{+}^2-\DET{-}^2 \right)^2+4|\Delta_{\pm}|^2\left[ \left( \DE{+}+\DE{-} \right)^2+|\Delta_+|^2+|\Delta_-|^2
                \right]-8\text{Re} \left( \Delta_{\pm}^2\Delta_+^*\Delta_-^* \right) }}\,, \label{eq:EigenvaluesFiniteB2}
\end{align}
where $E_{\pm}$ are the quasiparticle energies with $B=0$ (cf.  Appendix A for the derivation of the involved quantities).
\end{widetext}
 The  remaining dimensionless functions are 
\begin{align}
    \Dg{1}(\bold{k})   &= -\dfrac{2\DxiT{+}(\bold{k})}{\Delta_{1}(\bold{k})}            
                                    \DV{11}(\bold{k})\DV{41}^*(\bold{k}) \,,\\
    \Dh{1}(\bold{k})   &= -\dfrac{2\DxiT{+}(\bold{k})}{\Delta_{1}(\bold{k})}   
                                    \DV{21}(\bold{k})\DV{31}^*(\bold{k})  \,, \\
    \Dg{2}(\bold{k}) &= -\dfrac{2\DxiT{-}(\bold{k})}{\Delta_{2}(\bold{k})}              
                                    \DV{22}(\bold{k})\DV{32}^*(\bold{k}) \,, \\
    \Dh{2}(\bold{k}) &= -\dfrac{2\DxiT{-}(\bold{k})}{\Delta_{2}(\bold{k})}               
                                    \DV{12}(\bold{k})\DV{42}^*(\bold{k})\,.
\end{align}
The $\DV{}$-products in the above expressions are given by
\begin{align}
   \DV{11}\DV{41}^* = &\ a_{\bold{k}}^2 \DU{11}\DU{31}^* - b_{\bold{k}}^2 \DU{21}\DU{41}^* \notag\\
                       &+a_{\bold{k}}b_{\bold{k}}\left( \DU{21}\DU{31}^* - \DU{11}\DU{41}^* \right)\\
   \DV{21}\DV{31}^* = &\ b_{\bold{k}}^2 \DU{11}\DU{31}^* - a_{\bold{k}}^2 \DU{21}\DU{41}^* \notag\\
                       &-a_{\bold{k}}b_{\bold{k}}\left( \DU{21}\DU{31}^* - \DU{11}\DU{41}^* \right)\\
   \DV{22}\DV{32}^* = &\ b_{\bold{k}}^2 \DU{12}\DU{32}^* - a_{\bold{k}}^2 \DU{22}\DU{42}^* \notag\\
                       &-a_{\bold{k}}b_{\bold{k}}\left( \DU{22}\DU{32}^* - \DU{12}\DU{42}^* \right)\\
   \DV{12}\DV{42}^* = &\ a_{\bold{k}}^2 \DU{12}\DU{32}^* - b_{\bold{k}}^2 \DU{22}\DU{42}^* \notag\\
                       &+a_{\bold{k}}b_{\bold{k}}\left( \DU{22}\DU{32}^* - \DU{12}\DU{42}^* \right).
\end{align}
For simplicity, we left out the $\bold{k}$ dependence of the entries of the unitary transformations $\DV{},\DU{}$ in the above expression and dropped  the index $\tau=K$ in $a_{\bold{k}},b_{\bold{k}}$, defined in Eq. (\ref{eq:AnsatzDiagonalizationBField}), from now on.
%Finally, we conclude that the new gap equation acquires a very similar form as the one without the magnetic field in Eq. (\ref{eq:GapEquation}). However, we now have the new functions $\Dg{},\Dh{}$ which complicate matters by a lot. 
The explicit form of the functions $g_i$ and $h_i$  remains unknown due to the rather complicated expressions for the transformation in Eq. (\ref{eq:unitaryTrafo}).  However, this is not so dramatic for our purpose, as we will see later on. 

%Further notice that the magnetic field dependence is now incorporated within these new functions and also within the new Eigenenergies $\DxiT{i}$.
Before we proceed, let us observe that in the  case  $\bold{B}=0$  it holds $a_{\bold{k}}^2=1$ and  $b_{\bold{k}}^2=0$, and in turn $\Delta_+ = -\Delta_{1},\ \Delta_- = \Delta_{2}$ and $\Delta_{\pm} = 0$. In this limit the unitary transformation greatly simplifies,  we recover the Bogoliubov transformation from Eq. (\ref{eq:BogoliubovTrafo}) and we find $\Dg{i} = 1$,  $\Dh{i} = 0$.
\cite{noteBtoZero}
The functions $\tilde\chi_i(\bold{k})$ and their coefficients, given by the effective potentials in Eq. (\ref{eq:GapEqFiniteB}),  reduce to their much simpler form in Eq. (\ref{eq:GapEquation}) and we recover the zero field gap equation from the previous section.  

To address the case $\bold{B} \neq 0$ we first assume constant interaction potentials which again leads to $\bold{k}$-independent gaps. By defining the new quantities $\tilde\alpha_i = \frac{1}{N\Omega}\sum_{\bold{k}}g_i(\bold{k})\tilde\chi_i(\bold{k})$ and $\tilde\beta_i = \frac{1}{N\Omega}\sum_{\bold{k}}h_i(\bold{k})\tilde\chi_i(\bold{k})$ we can rewrite the gap equation as 
\begin{align}
    &0 = \mathbf{M}(B)\cdot \bold{\Delta} = \notag \\
    &\!\!\! \left[ \mathbf{1} + \!
		\begin{pmatrix}
			\Uintra \tilde\alpha_1-\Uinter \tilde\beta_1 && \Uintra \tilde\beta_2 -\Uinter \tilde\alpha_2 \\
			\Uintra \tilde\beta_1-\Uinter \tilde\alpha_1 && \Uintra \tilde\alpha_2-\Uinter \tilde\beta_2
		\end{pmatrix}
	\right]\!\!\!
	\begin{pmatrix}
	   \Delta_1\\
	   \Delta_2
	\end{pmatrix} .
	%\bold{\Delta}
		\label{eq:compactGapEqFiniteB}
\end{align}
The relation above  yields a modified condition for non-trivial solutions of the gap equation
\begin{align}
   & \det \tilde{\bold{M}}(B) = \UattrI\Uintra \tilde\alpha_1\tilde\alpha_2 + \Uintra(\tilde\alpha_1+\tilde\alpha_2)+1 \notag\\
                  &-\left[ \UattrI\Uintra \tilde\beta_1\tilde\beta_2 + \Uinter\left( \tilde\beta_1+\tilde\beta_2 \right) \right] = 0\,.
    \label{eq:modifiedDeterminant}
\end{align}
%
%Since we already know that $\Dg{i} \equiv 1$ and $\Dh{i} \equiv 0$ in the limit $\bold{B}=0$, we can also deduce that $\alpha_i$ simply reduces to the one from before and $\beta_i=0$. Hence, we are left with only the first row of the above equation which is the same we found before in Eq. (\ref{eq:determinant}).

We now turn to the determination of the critical magnetic field $B_c(T)$ at a given temperature $T$. The procedure  is  similar to the one we used to find the critical temperature. There, we considered a large enough temperature which closes both gaps, i.e. we set the gaps to zero, and used the condition from the determinant in Eq. (\ref{eq:determinant}) to obtain $T_c$. 
We will now consider a fixed temperature $T<T_c$ and will  look for the magnetic field $B_c(T)$ that
closes  both gaps. For this purpose, we use the magnetic field dependent determinant equation, Eq. (\ref{eq:modifiedDeterminant}). At the critical field  
the quasiparticle dispersions $\DxiT{n}(\bold{k})$  reduce to the eigenenergies $\DE{n}(\bold{k})$ found in Eq. (\ref{eq:BfieldEigenenergies}).
However, the treatment of the limit behavior of the functions $\Dg{i}$ and $\Dh{i}$ requires more attention when 
$\Delta_{1}, \Delta_2 \to 0$.  Due to the special form of the unitary transformation $U$ found in the 
Appendix, we cannot set both of them to zero simultaneously since this leads to diverging prefactors. What we can
do instead is treating $\Dg{1},\Dh{1}$ and $\Dg{2},\Dh{2}$ separately by first keeping one of the two gaps  finite 
and setting the other one to zero.
%What we get out of this is that the effective gaps introduced in Eqs.
%(\ref{eq:effGap12}) reduce to the still finite gap  modulo real prefactors that carry the magnetic
%field dependence. 
By this the entries $\DU{ij}$ of the unitary transformation can be written in such a way that the divergences cancel.\cite{noteGapToZero}
Some of the terms there are multiplied  by the still finite gap  will eventually drop out upon setting also this gap to zero. The remaining parts finally yield the quite compact expressions   
\begin{align}
  & g_{i=1,2}^0(\bold{k}):= \Dg{i=1,2}(\bold{k})\big|_{\Delta_1=\Delta_2=0} \\ & \!= 
                      a_{\bold{k}}^2 \left(   a_{\bold{k}}^2 +  b_{\bold{k}}^2                 
                         \dfrac{2\DE{\pm}(\bold{k})}{\DE{+}(\bold{k})+\DE{-}(\bold{k})}\right) \notag 
                 \\ &     = a_{\bold{k}}^2 \left(   1 \pm  b_{\bold{k}}^2                 
                         \dfrac{\DE{+}(\bold{k})-\DE{-}(\bold{k})}{\DE{+}(\bold{k})+\DE{-}(\bold{k})}\right) \,,\\
  & h_{i=1,2}^0(\bold{k}):= \Dh{i=1,2}(\bold{k}) \big|_{\Delta_1=\Delta_2=0} \\ &= 
                      a_{\bold{k}}^2 b_{\bold{k}}^2\left( 1-
                         \dfrac{2\DE{\pm}(\bold{k})}{\DE{+}(\bold{k})+\DE{-}(\bold{k})}\right) \notag 
                      = \mp  a_{\bold{k}}^2b_{\bold{k}}^2                 
                         \dfrac{\DE{+}(\bold{k})-\DE{-}(\bold{k})}{\DE{+}(\bold{k})+\DE{-}(\bold{k})}\,.
\end{align}
Notice that 
\begin{equation}
    \Dg{i}^0(\bold{k})+\Dh{i}^0(\bold{k}) = a_{\bold{k}}^2 \,.
\end{equation}
%From now on, these functions are always referred to as just $\Dg{i}$ and $\Dh{i}$ without explicitly stating that they are evaluated at $\Delta_\uparrow=\Delta_\downarrow=0$.
%They further fulfill the relation $ \Dg{i}(\bold{k})+\Dh{i}(\bold{k}) = a_{\bold{k}}^2$. 
%\textcolor{blue}{A similar relation can also be expected for the general functions $\Dg{i}$ and $\Dh{i}$ which is, however, not obvious due to the complicated $\DV{}$-products appearing in their definitions.}

In order to find the numerical values of $\tilde\alpha_i$ and $\tilde\beta_i$, we  turn the momentum sums into integrals over $\bold{k}$. In the case $\bold{B}=0$ the next step was to move the integration from the momentum to the energy space  since the functions. In the present case, the function    $\tilde\chi_i(\bold{k})$ only depends on one of the eigenenergies $\DE{n}(\bold{k})$, but the functions $g_i^0$ and $h_i^0$ depend on both  $\DE{+}(\bold{k})$ and $\DE{-}(\bold{k})$. %However, since now both of these energies appear in each integral, this gets more complicated
Hence it is now easier to directly calculate the momentum integrals by using polar coordinates. One finds
\begin{align}
   \tilde \alpha_{i=1,2} &= \dfrac{1}{(2\pi)^2}\int \dif \bold{k}\, \Dg{i}^0(\bold{k}) \tilde\chi_i(\bold{k}) \notag \\
             &= \dfrac{1}{2\pi}  
                \int_{k_i^{\text{min}}}^{k_i^{\text{max}}} \dif k\, 
                        %a_{\bold{k}}^2 \left(   a_{\bold{k}}^2 +  b_{\bold{k}}^2                 
                        % \dfrac{2\DE{i}(\bold{k})}{\DE{1}(\bold{k})+\DE{2}(\bold{k})}\right)
                        \Dg{i}^0(k) \dfrac{\tanh{\dfrac{\beta}{2}\DE{\pm}(k)}}{2\DE{\pm}(k)}\,,\\
   \tilde \alpha_{i=1,2}+\tilde\beta_{i=1,2}  &=  \sum\limits_\bold{k} a_{\bold{k}}^2 \dfrac{\tanh{\dfrac{\beta}{2}\DE{\pm}(\bold{k})}}{2\DE{\pm}(\bold{k})}                          \notag\\
                      &=    \dfrac{1}{2\pi} \int_{k_i^{\text{min}}}^{k_i^{\text{max}}} \dif k\,  a_{k}^2   
                            \dfrac{\tanh{\dfrac{\beta}{2}\DE{\pm}(k)}}{2\DE{\pm}(k)},
\end{align}
where we used the sum rule for the functions $\Dg{i}$ and $\Dh{i}$. It allowed us to express $\tilde\beta_i$ in terms of $\tilde\alpha_i$ and obtain a much simpler integral. The boundaries for the integral over $k = |\bold{k}|$ are the momentum cutoffs corresponding to $\pm \Lambda$ in the energy integrals from before. They are defined by $\xi_i(k_i^{\text{max/min}}) = \pm \Lambda$, with $\xi_i(\bold{k})$ in Eq. (\ref{eq:hyperbolic_dispersion_mu}).
%which yields 
%
%\begin{align}
%k_i^{\text{max/min}} = \frac{1}{\hbar v_{F,i}} \sqrt{(\tilde{\xi}_i \pm \Lambda)^2-m_i^2}.  
%\end{align}
%

\begin{figure}[ht]
	\begin{center}
		\includegraphics[width=\columnwidth,angle=0]{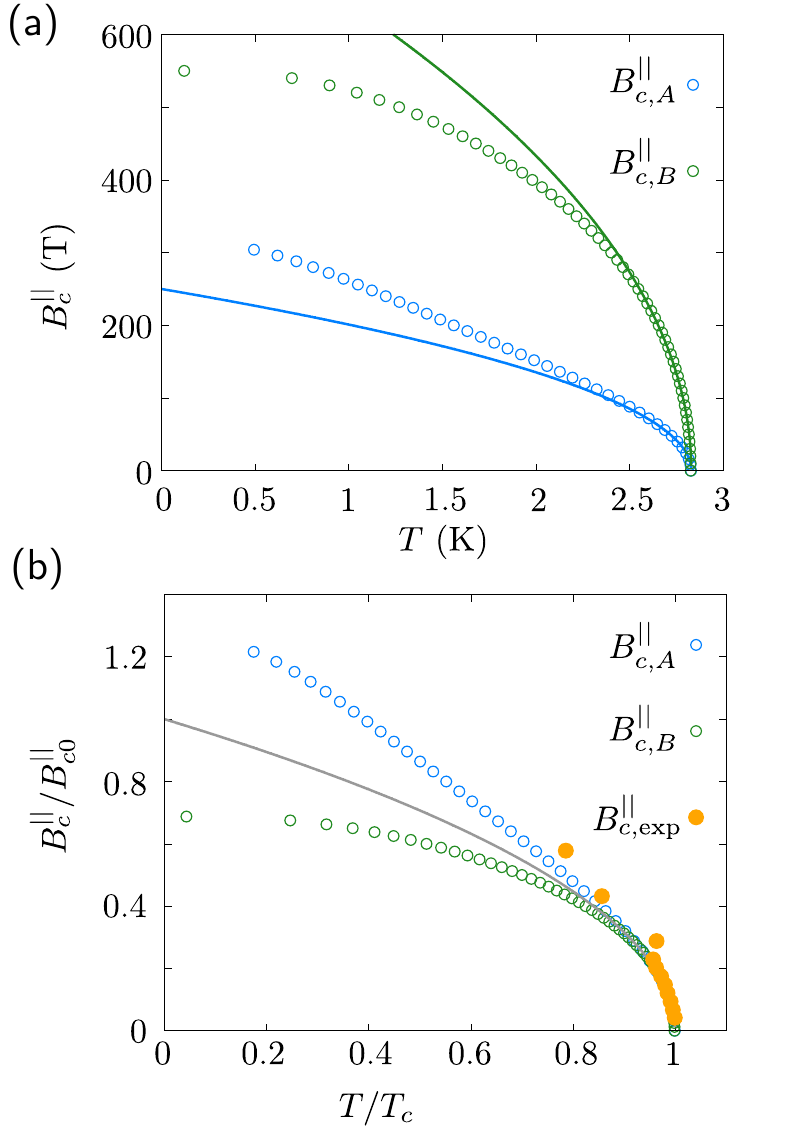}
		\caption{\small{Temperature dependence of the critical magnetic field. (a) Open circles are the numerical results, solid lines are the fits to $B_{c}^{\scriptsize{||}}(T) = B_{c}^{\scriptsize{||}}(0)\,\sqrt{1 - T/T_{c0}}$, where $T_{c0}$ is the critical temperature with $B_{\parallel} = 0$. The prefactors are $B_{c,A}^{\scriptsize{||}}(0) = 250$~T, $B_{c,B}^{\scriptsize{||}}(0) = 800$~T. (b) Rescaled results, where the grey line is $\sqrt{1-T/T_{c0}}$, and orange points are digitized data from Fig.~4 in Ref.~[\onlinecite{Xi:natphys2016}].  }}
		\label{fig:T-dependence-of-Bc}
	\end{center}
\end{figure}

Equipped with this, we are finally able to numerically solve the condition for non-trivial solutions defined in Eq. (\ref{eq:modifiedDeterminant}). The results of the simulations are shown in Fig. \ref{fig:T-dependence-of-Bc}, again for the $A$ and $B$ parametrizations, as well as  in comparison with the experimental data in Ref.~[\onlinecite{Wan:arxiv2021}]. 
On the one hand,  both parametrizations display the expected behavior 
\begin{align}
B_{c}^{\scriptsize{||}}(T) = B_{c}^{\scriptsize{||}}(0)\,\sqrt{1 - T/T_{c0}}
\end{align}
in the vicinity of the critical temperature at zero field, $T_{c0}$, in agreement with the experimental results \cite{Wan:arxiv2021}.  On the other hand, the two sets $A$ and $B$ differ qualitatively as the temperature is decreased. Within the $A$ parametrization  $B_{c}^{\scriptsize{||}}(T)$ displays almost a linear behavior at intermediate temperature, again in line with the experiment.  
The $B$ parametrization leads to much larger zero temperature critical fields than the $A$ one,  and  the scaled curve starts to deviate from the experimental data as the temperature decreases well below $T_c$. 
%\textcolor{blue}{The reason for this? Can we guess ?}
%\begin{figure}[ht]
%	\begin{center}
%		\includegraphics[width=8cm,angle=0]{figures/rescaled_criticalField.png}
%		\caption{\small{Temperature dependence of  the critical magnetic field: theory vs experiment.  }}
%		\label{fig:T-dependence-of-Bc-vs-exp}
%	\end{center}
%\end{figure}
%
%\begin{figure}[ht]
%	\begin{center}
%		%\includegraphics[width=5cm,angle=0]{figures/scheme_new.pdf}
%		\caption{\small{Magnetic field dependence of  the inner and outer gaps on an in-plane magnetic field.  }}
%		\label{fig:B-dependence-of-gaps}
%	\end{center}
%\end{figure}

\section{Conclusions} 
\label{Sec:Conclusions}
In summary, we have shown that two-bands superconductivity can  naturally arise from repulsive interactions in NbSe$_2$. At its origin is the fragmentation of the Fermi surface, which features hole pockets around the time-reversal related $K$,  $K'$ points, as well as around the  $\Gamma$ point. The latter, however, is not crucial for the effects discussed in this work where the superconductivity is driven by short-range intervalley scattering in combination with long-range intravalley interactions, for which two disjoint Fermi pockets are sufficient.  Superconductivity induced by repulsive
interactions which couple different valleys has been proposed
for the iron pnictides \cite{Mazin:nature2010}, bilayer graphene \cite{Guinea:prb2012}, heavily doped MoS$_2$ \cite{Roldan:prb2013}, and recently also for NbSe$_2$ \cite{Shaffer:prb2020}.  In our work we have extended in particular the treatment in Ref.~[\onlinecite{Roldan:prb2013}] to include  the effects of Ising SOC and  in-plane magnetic fields. For this purpose we have developed a fully microscopic approach based on a low energy, two-bands model for NbSe$_2$.  The minimal model fits  the hyperbolic dispersion for the holes near the $K$ and $K$' valleys to the outcomes of a three $d$-orbital tight-binding model for TMDCs. For that purpose, two different tight-binding  parametrizations for NbSe$_2$, called $A$ and $B$, are used. Due to the hyperbolic fit in combination with the Ising spin-orbit coupling, the two bands are gapped with two distinct gaps sharing the same critical temperature. As the temperature is decreased, the gaps at one valley acquire different size. Similar to conventional single gap BCS theory, we demonstrated a universal behavior of the average gap, when scaled by its zero temperature value, and with temperatures given in unit of the critical temperature.    The temperature  evolution of the scaled average gap  matches the data\cite{Wan:arxiv2021}, which are compatible with the presence of two distinct gaps. At finite magnetic field the two bands are also gapped. Due to the breaking of time-reversal symmetry, the quasiparticle spectrum acquires an intricate dependence on various pairing contributions with different character upon momentum inversion. The investigation of the different phases that our theory predicts at finite magnetic field will, in particular in relation to the possible observation of triplet superconductivity in [\onlinecite{Kuzmanovic:arxiv2021}], be the object of our future work. Here rather we focused on the dependence of the critical in-plane magnetic field on temperature. A quantitative agreement with the scaled data  in Ref.~[\onlinecite{Xi:natphys2016}] was found for one of the two chosen parametrizations, giving us confidence in the used microscopic low energy modeling. 

\section*{Acknowledgements}
We thank D. Kochan for useful discussions. M.M. thanks also C. Quay and D. Roditchev for their questions and suggestions.  The authors
 acknowledge financial support from the
 Elitenetzwerk Bayern via the IGK Topological Insulators and
 the Deutsche Forschungsgemeinschaft via the SFB 1277-II (subproject
 B04).

\newpage

\onecolumngrid

\newpage

\appendix

\section{ Diagonalization of the two bands superconducting Hamiltonian in finite in-plane magnetic field  }
 In the following we provide a way to diagonalize the BdG Hamiltonian in Eq. (\ref{eq:BdGMatrix}). Notice that in this appendix, for ease of notation, we use the subscripts $1,2$ instead of $+,-$. Also, we are going to introduce several new quantities whose expression is  listed at the very end of the section. 
Let us now consider the slightly more general case:
Given $\DE{1},\DE{2} \in \mathbf{R}$ and $\Delta_1, \Delta_2, \Delta_{12} \in \mathbf{C}$, we wish to diagonalize the following hermitian matrix,
\begin{align}
    \mathcal{H} = 
    \begin{pmatrix}
        \DE{1} & 0 & \Delta_1 & \Delta_{12} \\ 
        0 & \DE{2} & -\Delta_{12} & \Delta_2 \\
        \Delta_1^* & -\Delta^*_{12} & -\DE{1} & 0 \\
        \Delta^*_{12} & \Delta^*_2 & 0 & -\DE{2}
    \end{pmatrix}.
    \label{eq:AppendixH}
\end{align}
 I.e., we look for the eigenvalues and the unitary transformation that leads to the diagonal form of $\mathcal{H}$. 
We start by first dividing the matrix into the two parts
\begin{align}
    \mathcal{H} = \mathcal{H}_0 + \mathcal{H}_{12} = 
    \begin{pmatrix}
        \DE{1} & 0 & \Delta_1 & 0 \\ 
          0  & \DE{2} & 0 & \Delta_2 \\
        \Delta_1^* & 0 & -\DE{1} & 0\\
          0 & \Delta^*_2- & 0 & -\DE{2}
    \end{pmatrix}
    +
    \begin{pmatrix}
         &  &  & \Delta_{12} \\ 
          &  & -\Delta_{12} &  \\
         & -\Delta^*_{12} &  &  \\
        \Delta^*_{12} & &  & 
    \end{pmatrix} \,,
\end{align}
and proceed by diagonalizing just $\mathcal{H}_0$. Afterwards we express $\mathcal{H}_{12}$ in the   eigenbasis of $\mathcal{H}_{0}$ to get to the  matrix form of $\mathcal{H}$ in the new basis. Repeating this procedure eventually leads to a block diagonal matrix which can easily be diagonalized.
The unitary transformation that diagonalizes $\mathcal{H}_0$ is rendered as 
\begin{align}
    U_0 = 
    \begin{pmatrix}
         \Delta_1/\DXi{1} & 0 & -( \DET{1}-\DE{1} )/\DXi{1} & 0 \\
         0 & \Delta_2/\DXi{2} & 0 & -( \DET{2}-\DE{2} )/\DXi{2}  \\
         ( \DET{1}-\DE{1} )/\DXi{1} & 0 & \Delta_1^*/\DXi{1} & 0\\
         0 & ( \DET{2}-\DE{2} )/\DXi{2} & 0 & \Delta_2^*/\DXi{2}  \\
    \end{pmatrix}  \,,
    \label{eq:AppendixU0}
\end{align}
where the values of the entries fulfill $(\Delta_n/\DXi{n})^2 = \frac{1}{2}\left( 1+\DE{n}/\DET{n} \right)$ and 
$[( \DET{n}-\DE{n} )/\DXi{n}]^2 = \frac{1}{2}\left( 1-\DE{n}/\DET{n} \right)$. In the new basis, $\mathcal{H}$ reads
\begin{align}
    \tilde{\mathcal{H}} =
    U_0^\dagger \mathcal{H} U_0 = 
    \begin{pmatrix}
         \DET{1} & & & \\
         & \DET{2} & & \\
         & & -\DET{1} &\\
         & & & -\DET{2}\\
    \end{pmatrix}
    +
    \begin{pmatrix}
        0 & \DA & 0 & \DB    \\
         \DA^*& 0 & -\DB & 0 \\
        0 & -\DB^* & 0 & \DA^*\\
         \DB^* & 0 & \DA & 0  \\
    \end{pmatrix} \,.
%    =
%    \begin{pmatrix}
%         \DET{1} & \DA & & \\
%         \DA^*& \DET{2} & & \\
%         & & -\DET{1} &\DA^*\\
%         & & \DA & -\DET{2}\\
%    \end{pmatrix}
%    +
%    \begin{pmatrix}
%         & & & \DB     \\
%         & & -\DB &   \\
%         & -\DB^* & & \\
%        \DB^* & & &   \\
%    \end{pmatrix}
\end{align}
Notice that 
\begin{equation}
    |\DA|^2+|\DB|^2=|\Delta_{12}|^2\,, \quad  \DET{1}\DET{2}(|\DA|^2-|\DB|^2)=4\DE{1} \DE{2} |\Delta_{12}|^2 - \text{Re}(\Delta_{12}^2\Delta_1^*\Delta_2^*) \,.
    \label{eq:relations}
\end{equation}
Even though the total number of entries is unchanged, we are now left with only  two complex parameters $\DA$ and $\DB$ instead of the three provided by the gaps $\Delta_1$, $\Delta_2$ and $\Delta_{12}$ from before. In the second step we diagonalize the block diagonal part $\tilde{\mathcal{H}}_\DA$ of $\tilde{\mathcal{H}}$;  afterwards we express $\tilde{\mathcal{H}}_\DB$, the off-diagonal matrix which  contains the parameter $\DB$, in the new eigenbasis. Here, the corresponding unitary transformation is given by
\begin{align}
    U_1 = \dfrac{1}{\Da{1}}
    \begin{pmatrix}
         \DET{2}-G_{1} & -\DA & & \\
         -\DA^* & \DET{2}-G_{1} & &\\
         & & \DA^* & \DET{2}-G_{1} \\
         & & \DET{2}-G_{1} & \DA
    \end{pmatrix} \,,
\end{align}
and we have the relation $(\DET{2}-G_{1}) = -(\DET{1}-G_{2})$. In the new basis $\tilde{\mathcal{H}}$ looks like
\begin{align}
    K = U_1^\dagger \tilde{\mathcal{H}} U_1 = 
    \begin{pmatrix}
         G_{1} & & & \\
         & G_{2} & & \\
         & & -G_{2} &\\
         & & & -G_{1}\\
    \end{pmatrix}
    +
    \begin{pmatrix}
         & & \DR & \DS  \\
         & & \DT & -\DR \\
         \DR^*& \DT^* & &\\
         \DS^* & -\DR^* & & \\
    \end{pmatrix}\,,
\end{align}
which appears to be in a very similar form as the matrix $\mathcal{H}$ we have started with. Notice, however, that we have the important relations $|\DS|^2=|\DT|^2$ and $\DT \DR^* = \DS^* \DR$. The latter is the reason why we get a block diagonal matrix after a final rotation. For this, we first consider the off-diagonal contribution  $K_{\DR}$ of $K$ containing the $R$ elements.  One finds the transformation
\begin{align}
    U_2 = \dfrac{1}{\Da{2}}
    \begin{pmatrix}
         G_{2}+\tilde{G}_{1} & 0 & 0 & \DR\\
        0 & \DR & -(G_{1}+\tilde{G}_{2}) & 0 \\
         \DR^* & 0 & 0 & -(G_{1}+\tilde{G}_{2}) \\
        0 & G_{2}+\tilde{G}_{1} & \DR^* & 0 \\
    \end{pmatrix}
\end{align}
with the relation $G_{2}+\tilde{G}_{1} = G_{1}+\tilde{G}_{2}$. Performing the rotation now simply rearranges the entries $\DS$ and $\DT$ and we are left with the block diagonal matrix
\begin{align}
    \tilde{K} = U_2^\dagger K U_2 = 
    \begin{pmatrix}
        \tilde{G}_{1} & & & \\
        & -\tilde{G}_{1} & & \\
        & & \tilde{G}_{2} & \\
        & & & -\tilde{G}_{2}
    \end{pmatrix}
    +
    \begin{pmatrix}
        0 & \DS & & \\
        \DS^* & 0 & & \\
        & & 0 & \DT \\
        & & \DT^* & 0 \\   
    \end{pmatrix} \,,
\end{align}
whose diagonalization is trivial and can now be done within one step. The fourth and last transformation assumes the form
\begin{align}
    U_3 = 
    \begin{pmatrix}
        (\tilde{G}_{1}+\DxiT{1})/\Dp{1} & 0 & \DS/\Dp{1} & 0 \\
        \DS^*/\Dp{1} & 0 & -(\tilde{G}_{1}+\DxiT{1})/\Dp{1} & 0 \\
        0 & (\tilde{G}_{2}+\DxiT{2})/\Dp{2} & 0 & \DT/ \Dp{2} \\
        0 & \DT^*/\Dp{2} & 0 & -(\tilde{G}_{2}+\DxiT{2})/\Dp{2}
    \end{pmatrix}
\end{align}
and finally gives 
\begin{align}
    U_3^\dagger \tilde{K} U_3 = 
    \begin{pmatrix}
       \DxiT{1} & & & \\
       & \DxiT{2} & & \\
       & & -\DxiT{1} & \\
       & & & -\DxiT{2}
    \end{pmatrix}\,.
\end{align}
Ultimately, the full unitary transformation, which diagonalizes the matrix $\mathcal{H}$ in Eq. (\ref{eq:AppendixH}), can be explicitly calculated from the product of all $U$'s. We conclude the diagonalization with
\begin{align}
    U = U_0 U_1 U_2 U_3 = 
    \begin{pmatrix}
       %\color{blue} \DU{11} & \color{red}\DU{12} & \color{green}\DU{31}^* & -\DU{32}^* \\
       %\color{red}\DU{21} & \color{blue}\DU{22} & -\DU{41}^* & \color{green}\DU{42}^* \\
       %\color{green}\DU{31} & \DU{32} & \color{blue}-\DU{11}^* & \color{red}\DU{12}^* \\
       %\DU{41} & \color{green}\DU{42} & \color{red}\DU{21}^* & \color{blue}-\DU{22}^* \\
       \DU{11} & \DU{12} & \DU{31}^* & -\DU{32}^* \\
       \DU{21} & \DU{22} & -\DU{41}^* & \DU{42}^* \\
       \DU{31} & \DU{32} & -\DU{11}^* & \DU{12}^* \\
       \DU{41} & \DU{42} & \DU{21}^* & -\DU{22}^* \\
    \end{pmatrix}
    \label{eq:unitaryTrafo}
\end{align}
where each entry has a very similar counterpart
\begin{align*}
        \DXi{1}\Da{1}\Da{2}\Dp{1} \cdot \DU{11} &= 
            \Delta_1\left[ (\DET{2}-G_{1})(G_{2}+\tilde{G}_{1})(\tilde{G}_{1}+\DxiT{1})-\DA\DR\DS^* \right]
            -(\DET{1}-\DE{1})\left[ \DS^*(\DET{2}-G_{1})(G_{2}+\tilde{G}{1})+\DA^*\DR^*(\tilde{G}_{1}+\DxiT{1}) \right]\,, 
    \\
        \DXi{2}\Da{1}\Da{2}\Dp{2} \cdot \DU{22} &= 
            \Delta_2\left[ (\DET{2}-G_{1})(G_{2}+\tilde{G}_{1})(\tilde{G}_{2}+\DxiT{2})-\DA\DR\DS^* \right]
            -(\DET{2}-\DE{2})\left[ \DT^*(\DET{2}-G_{1})(G_{2}+\tilde{G}{1})+\DA\DR^*(\tilde{G}_{2}+\tilde{E}_{2}) \right]\,,
    \\
    \notag \\
        \DXi{1}\Da{1}\Da{2}\Dp{2} \cdot \DU{12} &= 
            \Delta_1\left[ \DA(G_{2}+\tilde{G}_{1})(\tilde{G}_{2}+\DxiT{2})+\DR\DT^*(\DET{2}-G_{1}) \right]
            -(\DET{1}-\DE{1})\left[ \DR^*(\DET{2}-G_{1})(\tilde{G}_{2}+\DxiT{2})-\DA\DS^*(G_{2}+\tilde{G}_{1}) \right]\,,
    \\
        -\DXi{2}\Da{1}\Da{2}\Dp{1} \cdot \DU{21} &= 
            \Delta_2\left[ \DA^*(G_{2}+\tilde{G}_{1})(\tilde{G}_{1}+\DxiT{1})+\DR\DS^*(\DET{2}-G_{1}) \right]
            -(\DET{2}-\DE{2})\left[ \DR^*(\DET{2}-G_{1})(\tilde{G}_{1}+\DxiT{1})-\DA\DS^*(G_{2}+\tilde{G}_{1}) \right]\,,
    \\
    \notag \\
    \DXi{1}\Da{1}\Da{2}\Dp{1} \cdot \DU{31} &= 
            \Delta_1^* \left[ \DS^*(\DET{2}-G_{1})(G_{2}+\tilde{G}_{1})+\DA^*\DR^*(\tilde{G}_{1}+\DxiT{1}) \right]
            +(\DET{1}-\DE{1})\left[ (\DET{2}-G_{1})(G_{2}+\tilde{G}_{1})(\tilde{G}_{1}+\DxiT{1})-\DA\DR\DS^* \right] \,
    \\
        \DXi{2}\Da{1}\Da{2}\Dp{2} \cdot \DU{42} &= 
            \Delta_2^* \left[ \DT^*(\DET{2}-G_{1})(G_{2}+\tilde{G}_{1})+\DA\DR^*(\tilde{G}_{2}+\DxiT{2}) \right]
            +(\DET{2}-\DE{2}) \left[ (\DET{2}-G_{1})(G_{2}+\tilde{G}_{1})(\tilde{G}_{2}+\DxiT{2})-\DA\DR\DS^* \right] \,
    \\
    \notag \\
        \DXi{1}\Da{1}\Da{2}\Dp{2} \cdot \DU{32} &= 
            \Delta_1^* \left[ \DR^*(\DET{2}-G_{1})(\tilde{G}_{2}+\DxiT{2})-\DA\DS^*(G_{2}+\tilde{G}_{1}) \right]
            +(\DET{1}-\DE{1}) \left[ \DA(G_{2}+\tilde{G}_{1})(\tilde{G}_{2}+\DxiT{2})+\DR\DT^*(\DET{2}-G_{1}) \right] \,,
    \\
        -\DXi{2}\Da{1}\Da{2}\Dp{1} \cdot \DU{41} &= 
            \Delta_2^* \left[ \DR^*(\DET{2}-G_{1})(\tilde{G}_{1}+\DxiT{1})-\DA\DS^*(G_{2}+\tilde{G}_{1}) \right]
            +(\DET{2}-\DE{2})\left[ \DA^*(G_{2}+\tilde{G}_{1})(\tilde{G}_{1}+\DxiT{1})+\DR\DS^*(\DET{2}-G_{1}) \right] \,,
\end{align*}
and where we used $\DA^*\DS = \DA\DT$. Further using the relations Eq. (\ref{eq:relations})  %$|\DA|^2+|\DB|^2=|\Delta_{12}|^2$ and $\DET{1}\DET{2}(|\DA|^2-|\DB|^2)=4\DE{1} \DE{2} |\Delta_{12}|^2 - \Re(\Delta_{12}^2\Delta_1^*\Delta_2^*)$ then 
leads to the explicit form of the eigenvalues. They read
\begin{align}
    \DxiT{n} &=  \dfrac{1}{\sqrt{2}}\sqrt{\DET{1}^2+\DET{2}^2+2|\Delta_{12}|^2 \pm 
                \sqrt{ \left( \DET{1}^2-\DET{2}^2 \right)^2+4|\Delta_{12}|^2\left[ \left( \DE{1}+\DE{2} \right)^2+|\Delta_1|^2+|\Delta_2|^2
                \right]-8\text{Re} \left( \Delta_{12}^2\Delta_1^*\Delta_2^* \right) }}\,. \label{eq:EigenvaluesFiniteB}
\end{align}
The transformation $U$, however, is still only given in an implicit form where each of our introduced abbreviations appear. 
%For our purpose, this form is sufficient since we are only interested in a special limit of this transformation.

The  list below shows the abbreviations introduced in the diagonalization process. Here, each block corresponds to the quantities that appear in one of the transformations $U_i$ and in the transformed matrix upon applying this transformation. 
\begin{alignat*}{3}
   %-----------------------------------------------------------------------------------------
    I. \quad && 
    \DET{n} &= \sqrt{\DE{n}^2+|\Delta_n|^2} &
    \DA     &= \dfrac{1}{\DXi{1}\DXi{2}}\left[ \Delta_{12}\Delta_1^*\left( \DET{2}-\DE{2} \right)  
                -\Delta_{12}^*\Delta_2\left( \DET{1}-\DE{1} \right) \right] \\
    %------
    &&
    \DXi{n} &= \sqrt{2\DET{n}\left( \DET{n}-\DE{n} \right)} & 
    \DB     &= \dfrac{1}{\DXi{1}\DXi{2}}\left[ \Delta_{12}\Delta_1^*\Delta_2^*            
                +\Delta_{12}^*\left( \DET{1}-\DE{1} \right) \left( \DET{2}-\DE{2} \right)\right]\\
    \\
    %-----------------------------------------------------------------------------------------
    II. \quad &&
    %\left.            
    %\begin{matrix}
    %    \Dchi{1}\\
    %    \Dchi{2}
    %\end{matrix}
    %\right\}
    G_{n}&= \dfrac{1}{2} \left[ \left( \DET{1}+\DET{2} \right) \pm \sqrt{\left( \DET{1}+\DET{2} \right)^2
                +4|\DA|^2} \right] & \quad
    \DR     &= \dfrac{1}{\Da{1}^2} \DB \left[ |\DA|^2 - \left( \DET{2}-G_{1} \right)^2 \right] \\
    %------
    &&
    \Da{1}  &= \sqrt{\left( \DET{2}-G_{1} \right)^2+|\DA|^2} &
    \DS     &= \dfrac{2}{\Da{1}^2} \DA \DB \left( \DET{1}-G_{1} \right) \\
    %------
    &&
    &&
    \DT     &= \dfrac{2}{\Da{1}^2} \DA^* \DB \left( \DET{2}-G_{1} \right) \\
    \\
    %-----------------------------------------------------------------------------------------
    III. \quad &&
    %\left.            
    %\begin{matrix}
    %    \DchiT{1}\\
    %    \DchiT{2}
    %\end{matrix}
    %\right\}
    \tilde{G}_{n} &= \pm\dfrac{1}{2} \left[ \left( G_{1}-G_{2} \right) \pm \sqrt{\left( G_{1}+G_{2}
                \right)^2 +4|\DR|^2} \right] &
    \Da{-}    &= \sqrt{\left( G_{2}-\tilde{G}_{1} \right)^2+|\DR|^2} \\
    \\
    %-----------------------------------------------------------------------------------------
    IV. \quad &&
    \DxiT{n}&= \sqrt{\tilde{G}_{n}^2+|\DS|^2} &
    \Dp{n}  &= \sqrt{\left( \tilde{G}_{n}+\DxiT{n} \right)^2+|\DS|^2}\\
    %-----------------------------------------------------------------------------------------
\end{alignat*}

\twocolumngrid

\bibliographystyle{apsrev}
\bibliography{references_two-gaps}

\begin{thebibliography}{40}
\bibitem[{not({\natexlab{a}})}]{noteTwoRoots}
\bibinfo{note}{{The polynomial yields two real solutions and hence also two
  different values for $T_c$. This other value can be obtained by simply
  changing the sign in front of the root. However, the resulting $T_c$ leads to
  a contradiction with the assumption $\Lambda \gg k_B T_c$ and is hence
  discarded.}}
\bibitem[{not({\natexlab{b}})}]{noteBtoZero}
\bibinfo{note}{{Since $\DA,\DB = 0$ in our diagonalization process, the
  Hamiltonian assumes a diagonal form already after the first step. All of the
  introduced eigenenergies there will then reduce to $\DET{i}$ which is simply
  the one from Eq. (\ref{eq:QuasiparticleDispersion}). Further, the $\DV{}$
  products appearing in the functions $\Dh{i}$ will vanish because every term
  in them is either multiplied by $\DA$ or $\DB$. The $\DV{}$ products in
  $\Dg{i}$ on the other side reduce to $-\Delta_{K\sigma}/2\DET{i}$.}}
\bibitem[{not({\natexlab{c}})}]{noteGapToZero}
\bibinfo{note}{{The problematic factors are the $\DXi{i}$ introduced in the
  diagonalization which have to be paired with either $|\Delta_n| \sim
  |\Delta_{K\sigma}|$ or $(\DET{n}-\DE{n})$. Setting then both
  $\Delta_{K\sigma}$ to zero will lead to a finite result (cf. Eq.
  (\ref{eq:AppendixU0})).}}
\end{thebibliography}
\end{document}